\documentclass[twocolumn]{aastex701}

\AfterEndPreamble{%
  \hypersetup{citecolor=blue}%
}
\usepackage{graphicx} 
\usepackage{subcaption}
\usepackage{appendix}
\usepackage[T1]{fontenc}
\usepackage{outlines}
\usepackage{color}
\usepackage{amsmath}    
\usepackage{amssymb}    
\usepackage{newtxtext,newtxmath}

\usepackage{threeparttable}
\usepackage{ulem}

\definecolor{blazeorange}{rgb}{1.0, 0.4, 0.0}
\definecolor{seagreen}{rgb}{0.18, 0.55, 0.34}
\definecolor{rufous}{rgb}{0.66, 0.11, 0.03}
\definecolor{royalfuchsia}{rgb}{0.79, 0.17, 0.57}
\definecolor{scarlet}{rgb}{1.0, 0.13, 0.0}
\definecolor{royalpurple}{rgb}{0.47, 0.32, 0.66}
\hypersetup{linkcolor=blue,citecolor=green,filecolor=cyan,urlcolor=magenta}

\newcommand{\lta}{\lower 2pt \hbox{$\, \buildrel {\scriptstyle <}\over {\scriptstyle \sim}\,$}}
\newcommand{\gta}{\lower 2pt \hbox{$\, \buildrel {\scriptstyle >}\over {\scriptstyle \sim}\,$}}

\def\MBH{M_{\rm BH}}
\def\Ms{M_{\star}}
\def\ms{m_{\star}}

\def\dNstar{dN_*/dM_*}
\def\dNBH{dN_{\rm BH}/dM_{\rm BH}}
\def\dRBH{d{\cal R}_{\rm BH}/dM_{\rm BH}}
\def\mbh6{M_{\rm BH,6}}
\def\L44{L_{\rm peak,44}}

\begin{document}

\title{
The TDE Population from First-Principles Models of Stellar Disruption and Debris Dynamics}

\author[0000-0002-7964-5420]{Tsvi Piran}
\affiliation{Racah Institute for Physics, The Hebrew University, Jerusalem, 91904, Israel}
\email{tsvi.piran@mail.huji.ac.il}

\author[0000-0002-2995-7717]{Julian Krolik}
\affiliation{Physics and Astronomy Department, Johns Hopkins University, Baltimore, MD 21218, USA}
\email{jhk@jhu.edu}
\correspondingauthor{Julian Krolik}

\author[0000-0003-2012-5217]{Taeho Ryu}
\affiliation{JILA, University of Colorado and National Institute of Standards and Technology, 440 UCB, Boulder, 80308 CO, USA}
\affiliation{Department of Astrophysical and Planetary Sciences, 391 UCB, Boulder, 80309 CO, USA}
\email{taeho.ryu@colorado.edu}
\affiliation{Max-Planck-Institut f\"ur Astrophysik, Karl-Schwarzschild-Str. 1, Garching, 85748, Germany}

\date{January 2026}

\begin{abstract}

We present a physically-grounded population model for optical tidal disruption events (TDEs) that combines first-principles hydrodynamic simulations of stellar disruption with statistical inference of the underlying stellar and black hole populations.
The model's 
{prediction of peak luminosity is based directly} on recent global simulations that follow the disruption self-consistently and
contains no tunable parameters related to the emission physics. We construct the predicted joint distribution of peak luminosity and black hole mass, including both full and partial disruptions, and compare it to a sample of observed TDEs using Bayesian inference and Markov chain Monte Carlo sampling. We find that the model reproduces the  distribution in
the ($\MBH,L_{\rm peak}$)  plane for the bulk of the observed TDE population with good statistical consistency.
The data strongly favor an old stellar population, with a sharp suppression of stars above $M_* \simeq 1.5 - 2\,M_\odot$.  They also indicate that, at fixed stellar mass, the volumetric TDE rate is nearly independent of black hole mass. Partial disruptions contribute a substantial fraction $(\sim30\%)$ of detected events in flux-limited samples and are essential for reproducing the observed distribution. The inferred population properties are robust to {different approximations to} the stellar mass–radius relation, although the {event rate at}
high-luminosity  is sensitive to  {the form of this relation for} massive stars.
We  predict a large population of {difficult to detect} low-luminosity TDEs, implying that the true volumetric TDE rate may exceed that inferred from present samples by up to an order of magnitude.
\end{abstract}

\section{Introduction}

In recent years, transient
flares from galactic nuclei having a wide range of properties have been found.  In some, the peak luminosity is in the range associated with Seyfert galaxies, $\sim 10^{43} - 10^{44}$~erg/s, lasts for a few months, and fades away 
with little change in color \citep{WeversRyu2023}.  These are generally identified as tidal disruption events (TDEs), in which a main-sequence star passes so close to a supermassive black hole (SMBH) that it is pulled apart tidally and loses all, or at least a large fraction, of its mass.  How exactly this results in a flare remains controversial \citep{Strubbe2009,Piran+2015,MetzgerStone2016,Krolik+2016,vanVelzen+2019a,Mummery-SB2020,Metzger2022,Mummery+2024,Krolik+2025}.
In a few cases, the flare may be repeated a year or two later, indicating only partial disruption \citep{WeversRyu2023,Broggi+2024,Somalwar+2025}, but in most, no recurrence has been seen over a decade of monitoring.

There are now so many known TDEs, or at least TDE candidates, that the aggregated data can be approached statistically \citep{Hammerstein+2023,Yao+2023}. The distributions of observed properties can then be compared to those predicted by specific dynamical models for TDEs that relate the stellar mass $M_*$ and the black hole mass $M_{\rm BH}$ to the peak luminosity $L_{\rm peak}$ and other observables.  All such predictions, however, rest on two assumed functions.  One is an assumed present-day stellar mass function for galactic nuclei $\dNstar$, which may or may not match the overall stellar mass function in the host galaxy.  The other is an assumed functional relation between the event rate per cosmological volume and the black hole mass $\dRBH$.  This rate is the product of the black hole mass function $\dNBH$ and a function describing how the rate of TDEs per black hole depends
on $M_{\rm BH}$. 

\begin{figure}
\includegraphics[width=0.95\linewidth]
{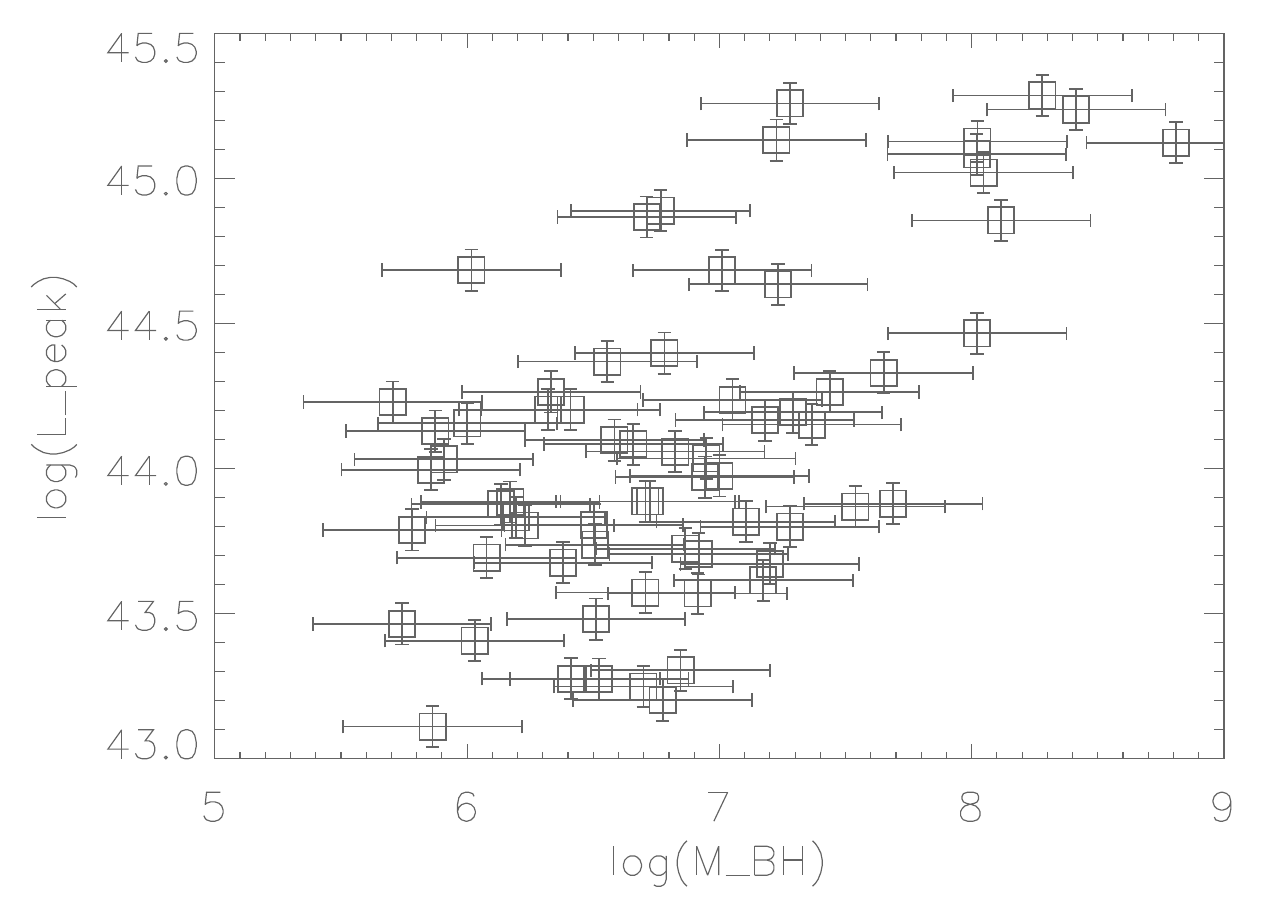}
\caption{The observed data points in the \texttt{ManyTDE} sample \citep{MvV}. $L_{\rm peak}$ is in erg~s$^{-1}$ and $M_{\rm BH}$ has units of $M_\odot$. As discussed in the text, events with $M_{\rm BH} \gtrsim 10^8 M_\odot$ must have dynamics qualitatively different from those involving lower-mass black holes. }
\label{fig:obsdata}
\end{figure}

The scatter in $L_{\rm peak}$ within the observed sample of events is so large (see Fig.~\ref{fig:obsdata}) that single-parameter ‘trends’ (e.g., as used in \cite{MvV2025})
fail to capture the demographics of the population, and correlation analyses of trends with $\MBH$ are sensitive to both the stellar mass and the black hole distributions. Meaningful comparison between observed TDE populations and theoretical models therefore requires analysis of the full two-dimensional distribution in the ($\MBH, L_{\rm peak}$) plane, together with explicit assumptions about the stellar mass function and the black hole event-rate distribution.

Here, we will make use of a 
recent advance in theoretical work on this subject: the concurrence in results of three very differently-executed global simulations of stellar tidal disruptions \citep{Ryu2023b,SteinbergStone2024,Price+2024}.  In all three, the event was studied in a self-consistent manner all the way from the star's approach to the black hole, through its disruption, and then through the much longer time over which its debris is deposited into an extended irregular and eccentric accretion flow.  {Using these results as a foundation  for quantitative calibration \citep{Krolik+2025}, we can predict the flare's peak luminosity as a function of $M_*$ and $M_{\rm BH}$ {\it with no free parameters}.   Combining this physical prediction with parameterized descriptions of $dN_*/dM_*$ and $d{\cal R}_{\rm BH}/dM_{\rm BH}$, we will explore the predicted population's sensitivity to these poorly-known functions.  We will then use a large observed sample to determine the forms of $dN_*/dM_*$ and $d{\cal R}_{\rm BH}/dM_{\rm BH}$ for which our predicted population is consistent with the observed population.}
Importantly, throughout this effort we will include the contribution of partial disruptions, which are unavoidable in any scenario, as well as full disruptions. 

\section{Method Details}

\subsection{Underlying model}\label{sec:model}

Contrasting the self-gravity of a small mass with the tidal gravity of a nearby larger mass leads to an order-of-magnitude estimate of the radius within which the smaller mass may suffer severe structural damage.  For the case of a star passing a black hole, it is
\begin{equation}
r_t \sim R_* (M_{\rm BH}/M_*)^{1/3},
\end{equation}
where $R_*$ is the star's radius.  {A single power-law fit to the solar-abundance middle-age main-sequence mass-radius relation gives $R_* = R_\odot (M_*/M_\odot)^{0.88}$ \citep{Ryu+2020b}.}
The critical orbital pericenter within which the star is wholly destroyed depends on the density profile of the undisturbed star, and therefore differs from $r_t$ by a factor $\sim O(1)$:
\begin{equation}
R_T = \Psi(M_*,M_{\rm BH}) \, r_t.
\end{equation}
General relativistic effects introduce black hole mass-dependence when $R_T \lta 50 r_g$, for $r_g \equiv GM_{\rm BH}/c^2$ \citep{Ryu+2020d}.
Because $R_T$ is almost independent of $M_*$ from $M_* \sim 0.3 - 3 M_\odot$, the strength of the relativistic effects at $R_T$ is also almost independent of $M_*$, and $\Psi$ can be factored into two pieces: $\Psi(M_*,M_{\rm BH}) = \Psi_*(M_*) \Psi_{\rm BH}(M_{\rm BH})$ \citep{Ryu+2020a}.

Similarly, the width of the debris' specific energy distribution can be approximated by
\begin{equation}
\Delta\epsilon \sim GM_{\rm BH} R_*/r_t^2,
\end{equation}
but this, too, is subject to order-unity corrections \citep{Ryu+2020a,Ryu+2020b,Ryu+2020d}:
\begin{equation}
\Delta E = \Xi(M_*,M_{\rm BH}) \, \Delta \epsilon .
\end{equation}
Much like the function $\Psi$, the correction factor $\Xi$ depends on $M_*$ through the star's internal density profile, and general relativistic effects create $M_{\rm BH}$-dependence.   For the same reasons as for $\Psi$, $\Xi$ can also be factored into $\Xi_*(M_*) \Xi_{\rm BH}(M_{\rm BH})$ \citep{Ryu+2020a}.
Although both ``correction" factors, $\Psi$ and $\Xi$, are of order unity, they are important in shaping the observed TDE distribution.

Not surprisingly, when the star's pericenter $r_p$ is $> R_T$, only a portion of the star's mass $\Delta M$ is removed, and its ratio to the star's initial mass declines rapidly with increasing $r_p/R_T$.  As found by \citet{Ryu+2020c},
\begin{equation}\label{eq:partialmass}
\Delta M/M_* \approx (r_p/R_T)^{-3}.
\end{equation}
Although the debris mass can be less than $M_*$, the orbital quantities $a_0$ and $t_0$ depend only on $M_{\rm BH}$ and $M_*$ because the most bound debris comes from the outer layers of the star, the same region where mass is lost in partial disruptions.

Partial disruptions must be included in any study of TDEs events because they are inescapable; in fact, as we will show shortly, the number of partial disruptions with peak luminosity $\gtrsim 0.1$ of the full disruption peak luminosity for the same combination of $M_*$ and $M_{\rm BH}$ is at least as large as the number of full disruptions. This holds true even for the full ``loss-cone" of stars with angular momentum small enough to pass within $R_T$ of the black hole is full.  Any depletion of the full-disruption loss-cone increases the ratio of partial disruptions to full disruptions.

Because the debris' orbital energy is a very small fraction of the potential energy at pericenter, i.e., $\Delta {\epsilon}/(GM_{\rm BH}/r_p) \sim (M_*/M_{\rm BH})^{1/3} \ll 1$, the characteristic scale of the debris orbits' semimajor axis is much larger than the pericenter:
\begin{equation}
a_0/r_p = \Xi^{-1} \, (M_{\rm BH}/M_*)^{1/3}.
\end{equation}
It immediately follows that the debris orbits are, at least immediately after the disruption, highly eccentric.

If the debris orbits all shared the same line of apsides, the only place where shocks could occur, and thereby dissipate orbital energy into heat, would be at the ``nozzle" near the pericenter, where the debris converges vertically \citep{EvansKochanek1989,Kochanek+1994}, but this is generically a rather weak shock \citep{Kochanek+1994,Shiokawa+2015,SteinbergStone2024,Ryu2023b,Price+2024,FritzHu+2025,Kubli+2025}.
However, there are two mechanisms by which the lines of apsides could be different for different debris parcels.  When the pericenter is sufficiently small, relativistic dynamics rotate the line of apsides \citep{Rees1988}, but the rotation angle is very small unless $r_p \lta 10 r_g$ (e.g., as given in \citet{Hobson+2006}).
In addition, the finite duration of the disruption itself means that debris leaves the star and is injected into its new orbit over a range of locations along the star's orbit, and the orientation of a mass parcel's line of apsides depends on its injection point \citep{Shiokawa+2015}.  So long as the spread in apsidal directions is small, the stream intersections occur closer to apocenter than pericenter; consequently, none of the resulting shocks is strong enough to dissipate a large fraction of the debris' orbital energy \citep{Shiokawa+2015,Dai+2015}.  Nonetheless, as confirmed by the three global simulations mentioned earlier \citep{SteinbergStone2024,Ryu2023b,Price+2024}, these shocks, supplemented by the nozzle shock, are strong enough to be the ultimate source of the heat powering the flare's light as first argued by \citet{Piran+2015}.

At the order of magnitude level, one then expects the rate at which heat is generated by the shocks to be the energy dissipated per unit mass ($\sim \Delta E$) times the mass of debris ($\sim \Delta M_*$: $\Delta M_* = M_*$ in a full disruption) per orbital period ($t_0 \sim (GM_{\rm BH}/a_0^3)^{1/2}$).  When the debris is not very optically thick, the heat can be quickly radiated as light, but if the optical depth is larger, the photon luminosity may emerge more slowly than on an orbital timescale.  Because the optical depth is proportional to the mass surface density,
it is $\propto (\Delta M/M_*)\Xi^2 M_{\rm BH}^{-2/3} M_*^{7/3}/R_*^2$; thus, slow cooling is associated particularly with smaller $M_{\rm BH}$. Combining these two mechanisms and employing the simulations' calibration, this model predicts
\begin{equation}\label{eq:Lpeak definition}
{L_{\rm peak}} = \frac{L_0}{1 + t_{\rm cool,0}/t_0},
\end{equation}
where
\begin{equation}\label{eq:L0R}
L_0 = 0.86 \times 10^{44} \Xi^{5/2} \mbh6^{-1/6} \frac{\Delta M}{M_*} \ms^{2.67} (R_*/R_\odot)^{-5/2}\hbox{~erg~s$^{-1}$}
\end{equation}
and
\begin{equation}
\frac{t_{\rm cool,0}}{t_0} = 0.26 \frac{f}{0.15}\frac{h}{r} \Xi^{5/2} \mbh6^{-7/6} (\kappa/\kappa_T) \frac{\Delta M}{M_*} m_*^{2.67} (R_*/R_\odot)^{-5/2}.
\label{Eq:tc_t0}
\end{equation}
Here $f$ is the fraction of all the bound debris mass that lies at a radius within $a_0$ at a time $t_0$ after the star's passage through pericenter, and $h/r$ is the orbiting debris' aspect ratio (our fiducial value of $h/r=0.3$).
 Details on the derivation of these expressions can be found in \citet{Krolik+2025}.

As remarked earlier (eqn.~\ref{eq:partialmass}), in a partial disruption the fractional mass removed from the star $\Delta M/M_* = x^{-3}$,  where $x\equiv  r_p/R_T > 1$.  This factor appears in eqn.~\ref{eq:L0R} because the luminosity is proportional to the debris mass participating. The cooling time, $t_{\rm cool}$, is reduced by the same factor because it, too, is proportional to the debris mass (eqn.~\ref{Eq:tc_t0}). With these small adjustments, we have an expression for the peak luminosity in partial disruptions,
\begin{equation}
L_{\rm peak}^{(partial)}
= \frac{L_0}{x^3 + (t_{\rm cool,0}/t_0)}.
\label{eq:partial}
\end{equation}

Substitution of the approximate relation $R_* \propto M_*^{0.88}$ \cite{Ryu+2020b} into Eqs. \ref{eq:L0R}-\ref{Eq:tc_t0} yields the peak luminosity formula of \citet{Krolik+2025}.
Although this single power-law fit
gives a fair approximation to theoretical calculations of $R_*(M_*)$ {for middle-aged main-sequence stars (MAMS)}, it can be improved upon.  In a more precise description, the slope of this relation is slightly steeper for $M_* \lesssim 1 M_\odot$ and significantly shallower for larger $M_*$ \citep{Torres+2010}.  Moreover, at this level of precision, one should properly speak of a family of relations, with variations having to do with stellar age and elemental abundances \citep{GZ1985,Torres+2010,FeidenChaboyer2012,Dotter2016}, and fitting the relation to a broken power-law has yielded parameters with some scatter (e.g., as described most recently in \citet{Camposeo2025}).  For our purposes, the principal effects of using a broken power-law are to make the radii of massive stars rather smaller than given by the single power-law fit, and the radii of low-mass stars slightly smaller.

\begin{figure}
\includegraphics[width=0.95\linewidth]
{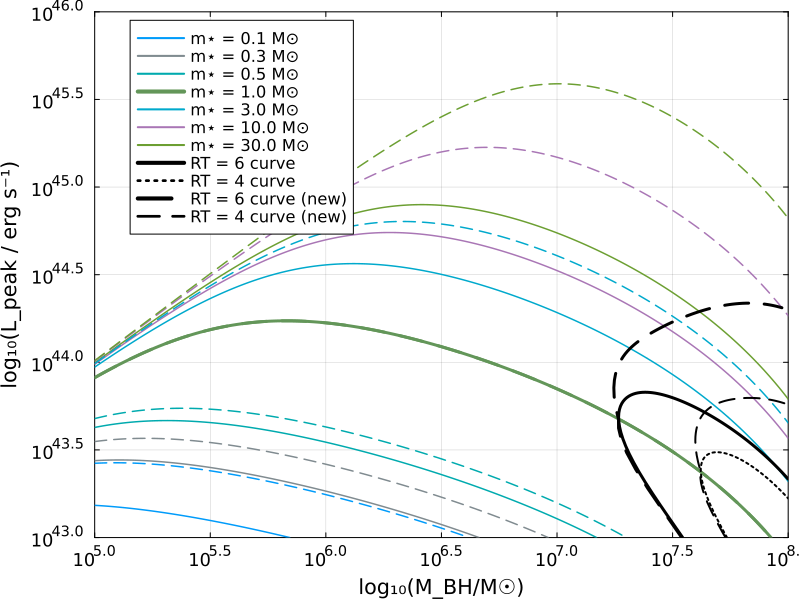}
\caption{Luminosities of full disruptions with stellar masses $M_*=0.1, 0.3, 0.5, 1, 3, 10, 30 M_\odot $ as functions of $M_{\rm BH}$.  Solid lines  are calculated using the single power-law mass-radius relation and dashed lines are calculated from the modified one. Also shown  are the curves of luminosity our model would predict for $R_T=6 r_g$ and $R_T = 4r_g$; extreme TDEs occur when $4r_g < r_p < 6r_g$ \citep{Ryu+2023}; direct capture ensues when $r_p < 4 r_g$ (an exact result for Schwarzschild spacetime, the angle-averaged result for Kerr \citep{Krolik+2020}). 
The luminosities of partial disruptions of a given $M_*$ form a family lying below the curve for a full disruption of such a star.}
\label{fig:Lpeak}
\end{figure}

Because both $L_0$ and $t_{\rm cool,0}/t_0$ are $\propto R_*^{-5/2}$, the peak luminosity can be very sensitive to the specific value of $R_*$.  For $L_0$, this dependence comes from the product of the potential depth ($\propto a_0^{-1} \propto R_*^{-1}$) and the orbital frequency ($\propto a_0^{-3/2} \propto R_*^{-3/2}$).  For the timescale ratio, this scaling results from the product of the diffusion time ($\propto a_0^{-1} \propto R_*^{-1}$) with the orbital frequency.  If, as is generally the case, the main-sequence mass-radius relation becomes shallower toward higher $M_*$, the resulting decrease in estimated $R_*$ can lead to a large increase in peak luminosity
(see Fig.~\ref{fig:Lpeak}).  As we will show later, this sensitivity to $R_*(M_*)$ can noticeably change the statistical behavior of the observed TDE population at the highest $L_{\rm peak}$.

When we wish to display the consequences of a refined mass-radius relation, we will employ a version of the broken power-law description that is roughly consistent with many of those published {(e.g., as in \citet{Torres+2010} and \citet{Camposeo2025})}:
\begin{equation}
R_*(M_*) = \begin{cases} (M_*/M_\odot) R_\odot & M_* \leq M_\odot \\
(M_*/M_\odot)^{0.6} R_\odot & M_* > M_\odot.\\
\end{cases}
\end{equation}
This form of the mass-radius relation implies that the $m_*$-dependence of both $L_0$ and $t_{\rm cool,0}/t_0$ becomes
\begin{equation}
L_0, t_{\rm cool,0}/t_0 \propto
\begin{cases} m_*^{0.17} & M_* \leq M_\odot \\
m_*^{1.17} & M_* > M_\odot\\
\end{cases}
\end{equation}
rather than $\propto m_*^{0.47}$.
Thus, the peak luminosity rises more slowly with $m_*$ for low-mass stars and more rapidly with $m_*$ for high-mass stars.
Note that because our form for $R_*(M_*)$ is keyed to the current radius of the Sun, it is automatically a description of the middle-age main-sequence mass-radius relation.

Some of the implications of our prescription for $L_{\rm peak}$ are illustrated in Fig.~\ref{fig:Lpeak}.  At low $M_{\rm BH}$, the peak luminosity for the full disruption of a star with mass $M_* \gtrsim M_\odot$ is Eddington-limited, so $L_{\rm peak} \propto M_{\rm BH}$.  It reaches a maximum at intermediate black hole mass, and then decreases at high $M_{\rm BH}$ because the timescale for the event increases, but, for a given $M_*$, the energy available for radiation is the same.  For stars of lower mass, the peak luminosity decreases slowly, but monotonically, with increasing $M_{\rm BH}$; they do not reach the Eddington limit even for small black holes, and are subject to the same increase in flare-duration time with $M_{\rm BH}$ as for higher-mass stars. Note, however, that the peak luminosity for high-mass stars is much more sensitive to introduction of a more precise mass-radius relation than it is for low-mass stars.  For a $0.3 M_\odot$ star, the highest peak luminosity for any black hole mass rises by only $\simeq 40\%$, but for a $30 M_\odot$ star it rises by a factor $\simeq 5$.

Our model is framed so as to apply to the majority of TDEs in a volume-limited sense, but it may not be appropriate in strongly relativistic events.   When the pericenter passage is very close to the black hole ($r_p \lesssim 10 r_g$), a number of relativistic mechanisms become important.   Apsidal precession could be large enough to raise the debris shock speeds, thereby substantially increasing the associated dissipation \citep{Rees1988}.  On the other hand,
the after-effects of passage through the nozzle shock may mitigate the total heating in the shocks created by apsidal precession  (H.S. Chan et al., in preparation). Black hole spin could drive orbital plane precession in the debris orbits, possibly delaying their intersection for a sizable time \citep{GuillochonRR2015}, or, at the very least, prevent confinement of the debris to a single plane.  In the latter case, any new features may depend strongly upon both the magnitude of the black hole spin and its direction relative to the star's orbital axis.  When $r_p \lesssim 6 r_g$, the star goes around the black hole multiple times, shedding mass all the while; this leads to drastically different dynamics \citep{Ryu+2023}.
When $r_p \lesssim 4r_g$, the entire star crosses the event horizon before it is tidally disrupted, suppressing any flare.

All of these mechanisms could lead to substantial changes in the relation between $M_*$, $M_{\rm BH}$, and $L_{\rm peak}$. 
One in particular deserves special emphasis.  When $M_{\rm BH} \gtrsim 10^8 M_\odot$, the critical pericenter for a full disruption becomes smaller than the critical radius for an extreme TDE if $M_* \lesssim 3 M_{\rm BH,8}^{1.2}$ assuming $R_* \propto M_*^{0.88}$; a broken power-law mass-radius relation would extend the range for this statement up to $\simeq 10M_\odot$.
In other words, when $M_{\rm BH} \gtrsim 1 \times 10^8 M_\odot$, ordinary disruptions simply do not occur for any stars but the most massive.  For this reason, it is not only our model that is inapplicable for $M_{\rm BH} \gtrsim 10^8M_\odot$---it is {\it all} models for the common variety of TDEs because the fundamental character of the event changes dramatically when $r_p \lesssim 6r_g$.

The second extreme limit is very high luminosity.  As just noted, if our model is applied using a single power-law mass-radius relation, it is unable to produce events with luminosity $\gtrsim 1 \times 10^{45}$~erg~s$^{-1}$.
Any events exhibiting higher luminosity would then have to be produced by a mechanism that differs in some substantial way from that of ordinary TDEs.
Therefore, when considering the single power-law mass-radius model, we impose a maximum luminosity of $10^{45}$~erg~s$^{-1}$ because that lies just above the maximum $L_{\rm peak}$ for a $30M_\odot$ star (see the top curve in Fig.~\ref{fig:Lpeak}).
However, the more accurate broken power-law relation lifts that limit to $\gtrsim 4 \times 10^{45}$~erg~s$^{-1}$, high enough to accommodate a number of the highest luminosity events.  In this case, no luminosity cap is required.

\subsection{Calculating the predicted event-rate population}

TDE events are often characterized by two quantities, peak luminosity and black hole mass.  The former comes directly from event observations; 
for the purpose of discriminating between TDE models, it is best to estimate the latter in ways independent of the models, such as methods based on
host galaxy properties and established correlations between them and the mass of their nuclear black hole. 
Here we will use the model just described to calculate the rate of events as a function of these two quantities; this prediction can then be compared to the statistics of detected events. 

One way to accomplish this calculation is to use analytic methods to reduce it to numerical integrations.  This method has the virtues of providing probability densities throughout the domain of interest and offering analytic insight into scalings.
When convenient, we will also deploy a Monte Carlo method, which has different advantages.

In both methods, we will assume that the mass-distribution of the stars subject to disruption is proportional to the stellar mass function $\dNstar$.  That the event rate per star should be roughly independent of $M_*$ is suggested by the fact that the critical pericenter for a complete disruption is approximately constant for $0.1M_\odot \lesssim M_* \lesssim 3 M_\odot$ \citep{Ryu+2020b}.  We define the distribution of black hole masses participating as $\dNBH$, but note that the rate per black hole is $\dRBH$, which combines $\dNBH$ with information about any physical effects associated with $M_{\rm BH}$ that might promote or suppress tidal disruptions (e.g., the larger physical cross section, or effects on the velocity distribution of nearby stars, etc.)

\subsubsection{Probability density}
Following the custom of the field, we will express the probability density in terms of $\log M_{\rm BH}$ and $\log L$.
The probability density {per volume}
for full disruptions with peak luminosity $L_{\rm peak}$ and black hole mass $M_{\rm BH}$ is then
\begin{multline}
\frac{\partial^2 N}{\partial \log M_{\rm BH}\partial \log L} =  \ln(10)\int dM_*  \delta \left[L - L_{\rm peak}(M_{\rm BH},M_*)\right]  \\
\times  L_{\rm peak}
\frac{\partial N}{\partial M_*} \frac{\partial {\cal R}_{\rm BH}}{\partial \log M_{\rm BH}} 
\left[1 - \frac{J_{\rm dc}^2}{J^2(R_T)} \right],
\end{multline}
where the pre-factor $\ln(10)$ is required to convert $\partial/\partial \ln L$  to $\partial/\partial \log L$.
The last factor in the integrand is the reduction in the full disruption rate due to the portion of the cross section occupied by direct captures, for which the critical angular momentum $J_{\rm dc} = 4$.  This factor could also be chosen so as to eliminate extreme TDEs ($4 r_g < r_p < 6 r_g$) as well, or account for a partially-filled loss-cone.

To evaluate the $\delta$-function, we change its argument to $M_*$.
We then have
\begin{multline}\label{eq:fullrates}
\frac{\partial^2 N} {\partial \log M_{\rm BH}\partial \log L} = 
\left[1 - \frac{J_{\rm dc}^2}{J^2(R_T)} \right] 
\frac{\partial {\cal R}_{\rm BH}}{\partial \log M_{\rm BH}} \, \\ \times \frac{\partial N}{\partial \log M_*^\prime}
\Big/ \,\frac{\partial\ln L_{\rm peak}}{\partial\ln M_*^\prime},
\end{multline} 
where the physical tidal radius, the stellar mass function, and the logarithmic derivative of peak luminosity by stellar mass are all evaluated at $M_*^\prime$, defined by the value of $M_*$ satisfying equation~\ref{eq:Lpeak definition} for $M_{\rm BH}$ and $L$.

For the volumetric probability density of partial disruptions, the procedure follows a similar approach, but with a few differences stemming from the fact that, for any $(M_{\rm BH},M_*)$ pair, there is a range of possible pericenters, from $r_p=R_T$ outward, and there is a corresponding decrease in the debris mass as $r_p$ increases beyond $R_T$.  Consequently, the ratio $x \equiv r_p/R_T$ must be integrated over.
By definition, $x > 1$ for all partial disruptions; we can then write the event rate for a given $L$ as
\begin{multline}
\frac{\partial^2 n}{\partial \log M_{\rm BH} \partial \log L} = \ln(10) \int \, d\log M_* \, \frac{\partial N}{\partial \log M_*} \frac{\partial {\cal R}_{\rm BH}}{\partial \log M_{\rm BH}} \\ 
\times \int_1\, dx \,
\frac{dP}{dx} L \delta \left[L - L_{\rm peak}^{(partial)}(M_{\rm BH},M_*,x)\right].
\end{multline}
Here we use $n$ for the partial disruption rate in contrast to $N$, the full disruption rate. 

The Newtonian probability density for events with pericenter $x R_T$ is uniform with respect to $x$ because the rate of events within a given $r_p$ is $\propto r_p$.
This is an excellent approximation to the Schwarzschild spacetime relativistic probability density for $r_p > 10r_g$,
a range in which most partial disruptions are found,
so we set $dP/dx = 1$.

Once again, we can change the argument of the $\delta$-function, this time from $L$ to $x$:
\begin{multline}
\frac{\partial^2 n}{\partial \log M_{\rm BH} \partial \log L} = \ln(10) \int \, d\log M_* \,  \frac{\partial N}{\partial \log M_*} \frac{\partial {\cal R}_{\rm BH}}{\partial \log M_{\rm BH}} \\
\times \int_1 \, dx \,
L \delta \left[x - \left(\frac{L_0}{L} - \frac{t_{\rm cool,0}}{t_0}\right)^{1/3} \right]\, \Big/ \, \partial L/\partial x.
\end{multline}

Because we know the analytic relation between $L$, $L_{\rm peak}^{(partial)}$, and $x$, it is easy to evaluate the last partial derivative: $\partial L/\partial x = -3x^2 L^2/L_0$.
We then find
\begin{multline}\label{eq:partialrates}
\frac{\partial^2 n}{\partial \log M_{\rm BH} \partial \log L} = (\ln(10)/3)\frac{\partial {\cal R}_{\rm BH}}{\partial \log M_{\rm BH}} \int \, d\log M_* \,  \\ \times \frac{\partial N}{\partial \log M_*} \frac{L_0/L}{\left(L_0/L - t_{\rm cool,0}/t_0\right)^{2/3}} .
\end{multline}
In other words, per star of a given mass, the density per unit volume of partial disruption events and unit $\log L$ is $\propto L^{-1/3}$ in the limit of rapid cooling, a relatively slow scaling.  Comparison of Eqns.~\ref{eq:fullrates} and \ref{eq:partialrates} immediately  shows that the rates of the two are comparable.

Although in principle it might be possible for the values of $M_*$ at which $x(M_*) > x_{\rm min}$ to be found in disconnected regions, in practice we find that the constraint $x(M_*) > x_{\rm min}$ translates directly to $M_* > M_{*,\rm min}$.   In the simplest case, $M_{*,\rm min}$ is the stellar mass at which the required luminosity is the peak luminosity of a full disruption, so the range of relevant stellar masses extends upward from that mass.

\section{Qualitative Properties of our Population Model}

\subsection{Parameterized mass function models}

To explore how our model reflects both $dN_*/dM_*$ and $d{\cal R}_{\rm BH}/dM_{\rm BH}$ (the latter subsuming $M_{\rm BH}$-dependent physical effects as well as the black hole population), we have examined the effects of broken power-laws for the stellar mass function and pure power-laws in $M_{\rm BH}$ for the black hole event rate.  The latter scaling is defined as $\dRBH \propto M_{\rm BH}^{-\alpha_{\rm BH}}$.
For initial qualitative studies, we have varied $\alpha_{\rm BH}$ from -2 to +2,
but as will be seen, only the range -0.5 to +0.5 is likely to be relevant to the observed population.
This power-law is defined for $10^5M_\odot \, \leq M_{\rm BH} \leq \, 10^8 M_\odot$.

To define the stellar mass function, we suppose that it is $\propto M_*^{-\alpha_1}$ up to mass $M_b$ and $\propto M_*^{-\alpha_2}$ for $M_* > M_b$.
This form invokes the least number of free parameters needed to describe commonly-observed stellar populations in many astronomical environments, including galactic nuclei as well as many other locales.
We have explored the consequences of varying these parameters over the ranges $0.5 \leq M_b/M_\odot \leq 4$, $0.5 \leq \alpha_1 \leq 3$, and $2 \leq \alpha_2 \leq 7$.  Formally, we consider a span of $M_*$ from $0.1 - 30M_\odot$, but with $\alpha_1 \leq 3$, stars with $M_* \lesssim 0.3 M_\odot$ contribute very few events in the observed range of luminosities.

\begin{figure*}
\hskip 1cm
\includegraphics[width=0.45\linewidth]
{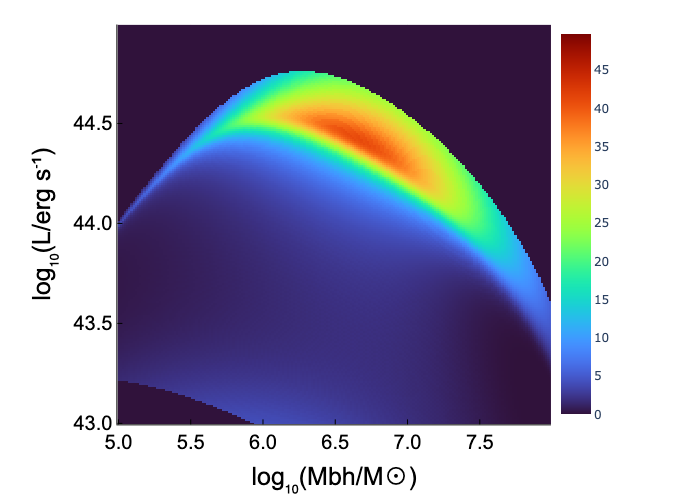}
\includegraphics[width=0.45\linewidth]{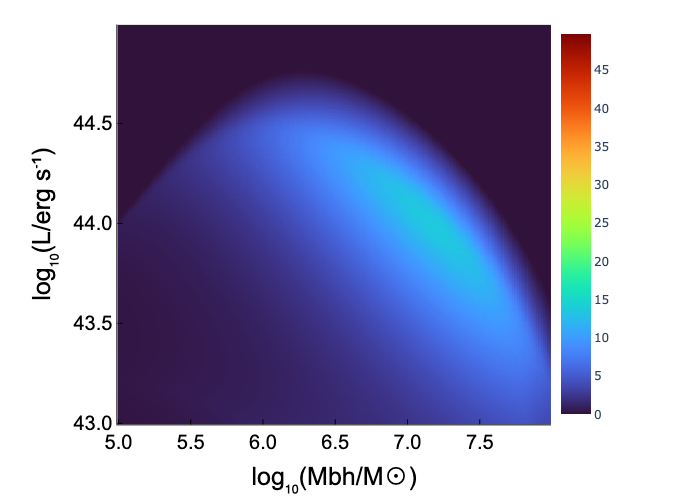}
\caption{The event rate density in the $\log M_{\rm BH} \times \log L_{\rm peak}$ plane for flux-limited samples: full disruptions (left) and partial disruptions (right).
In both cases, the stellar mass function's shape is the Salpeter IMF, and $\alpha_{\rm BH} = 0.$
Rates are shown on a linear scale calibrated by the colorbar; these scales are the same to make it easy to gauge the ratio of partial events to full.
}
\label{fig:fullvpartial}

\end{figure*}

Although there is strong evidence that the mass function in the Galaxy has the Kroupa form ($\alpha_1 \simeq -1.3$, $M_b \simeq 1 - 4$, $\alpha_2 \simeq -2.3$) \citep{Bastian+2010,delAlcazar+2025}, there are also indications that the present-day mass function in early-type galaxies, especially in their central regions, is closer to the Salpeter form ($\alpha_1 = \alpha_2 = -2.35$) \citep{Smith2020,vanDConroy2024}.  There are also hints that the steeper mass function in the centers of early-type galaxies is particularly associated with galaxies hosting larger black holes, $\gtrsim 1 \times 10^7 M_\odot$ \citep{vanDConroy2024}.  In addition, there is some evidence  there has been very little recent star formation in TDE hosts \citep{Zhang+2025}. For TDEs, we are, of course, especially concerned with the stellar mass function very close to the host galaxy's center, so we will treat its shape as an unknown, to be constrained by TDE data.

\subsection{General Trends}

The most direct way to illustrate how the predicted population depends on either the definition of our model or the values of its parameters 
is to present matched sets of figures.  Our primary measure of the population is the distribution of relative detected event rates as a function of $M_{\rm BH}$ and the bolometric peak luminosity $L_{\rm peak}$.  Although the formalism of the previous subsection yields the rates per unit volume, observational samples are nearly always flux-limited rather than volume-limited; to mimic the properties of flux-limited samples, we will multiply each volumetric density by $L^{3/2}$ to allow for the greater accessible volume afforded by greater luminosity.

\begin{figure*}
\hskip 1cm
\includegraphics[width=0.45\linewidth]
{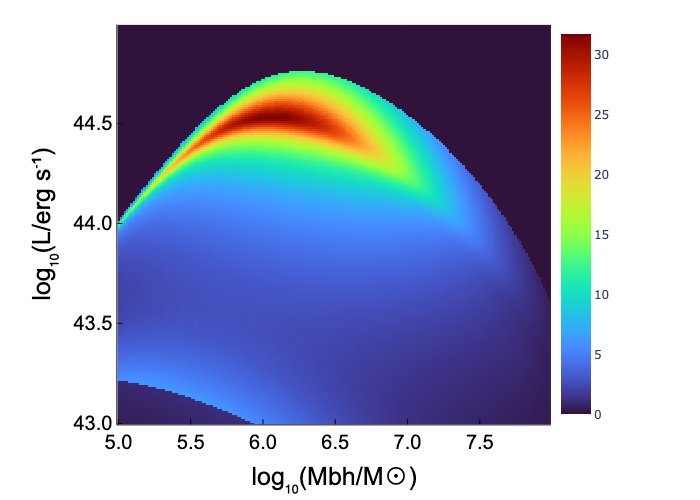}
\includegraphics[width=0.45\linewidth]
{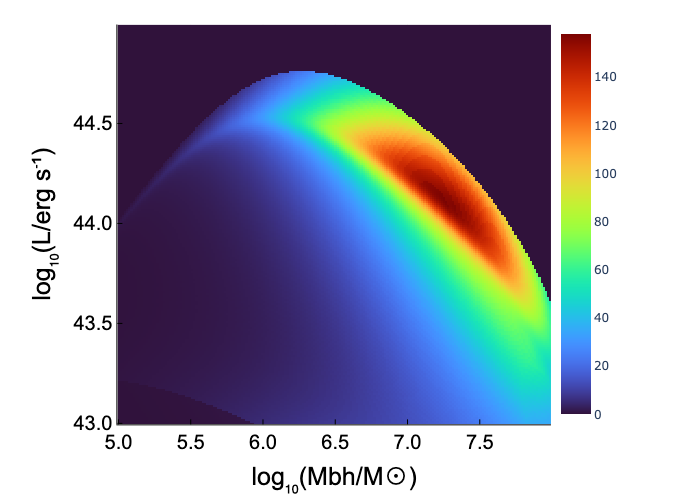}
\caption{The total (full + partial) event rate density in the $\log M_{\rm BH} \times \log L_{\rm peak}$ plane for flux-limited samples.  Both panels assume that the stellar mass function is proportional to a Salpeter IMF, but in the left panel $\alpha_{\rm BH} = -0.5$ and in the right panel $\alpha_{\rm BH} = +0.5$.
Rates (in arbitrary units) are shown on a linear scale calibrated by the colorbar.}
\label{fig:dRBH}
\end{figure*}

\subsubsection{Contribution of partial disruptions}

We begin with the relative contributions of full and partial disruptions.  As Fig.~\ref{fig:fullvpartial} makes plain, full disruptions alone are
strongly concentrated near the maximum possible luminosity (i.e., in the range $\sim 1 - 3 \times 10^{44}$~erg~s$^{-1}$) and only rarely yield smaller peak luminosity.
Partial disruptions, not surprisingly, in general produce lower luminosities, but they are numerous. In a volume-limited sense, when the stellar loss-cone is full, those with pericenters within twice the critical pericenter for full disruption occur at a rate comparable to the full disruption rate; to the degree the loss-cone is not full, there are more partial than full disruptions. Our predictions, both for these qualitative explorations and for our later likelihood analysis, will assume full loss-cones, so their partial fractions are lower bounds.
In addition, partial disruptions make up an especially large fraction of events at higher $M_{\rm BH}$ because more of the cross section for events with $r_p \leq R_T$ is occupied by direct captures.  For all these reasons, they are important to a different portion of the 
($M_{\rm BH},L_{\rm peak})$
phase space than full disruptions.

However, because they are usually less luminous than full disruptions, they suffer from being visible in only a limited volume, and are therefore less likely to appear in a flux-limited sample.  In rough terms, a flux-limited sample's ratio of the total number of partial disruptions  to the total number of full disruptions in the luminosity range of observed events is $\sim 0.3 - 0.4$.
Although the detailed shape of the event distribution combining partial with full disruptions depends on the stellar and black hole event rate distributions, it is clear that partial disruptions should always be included in predictions of the population.  In fact, there are observational indications suggesting there may be many unrecognized partial disruptions within current TDE samples \citep{Makrygianni+2025}.

\begin{figure*}
\hskip 1cm
\includegraphics[width=0.45\linewidth]{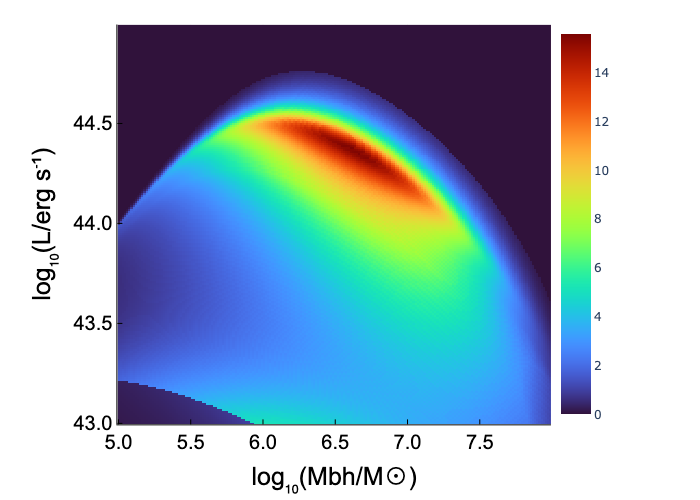}
\includegraphics[width=0.45\linewidth]{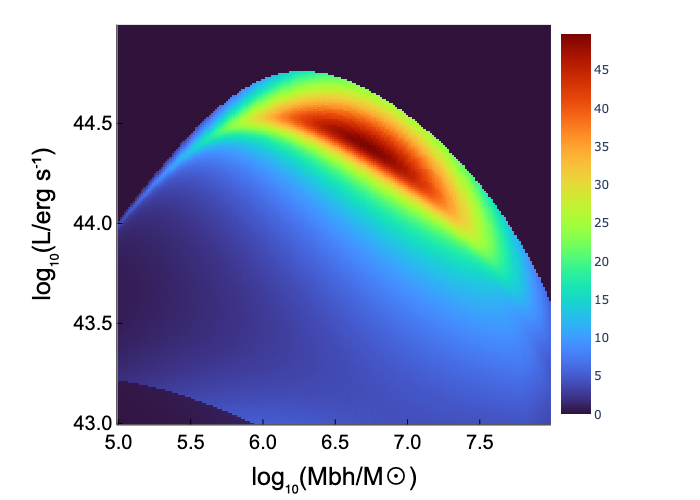}
\caption{The total (full + partial) event rate density in the $\log M_{\rm BH} \times \log L_{\rm peak}$ plane for flux-limited samples.
Both plots have $\alpha_{\rm BH} = 0$, but the left utilizes a Salpeter IMF mass function, cut-off at $1M_\odot$, i.e., $\alpha_1 = 2.35$, $\alpha_2 = 4$, and $M_b = 1M_\odot$, whereas the right has no cut-off: $\alpha_1 = \alpha_2 = 2.35$.
Rates (arbitrary units) are shown on a linear scale calibrated by the colorbar.}
\label{fig:dNstar}
\end{figure*}

\subsubsection{Effect of varying $\dRBH$}

Not surprisingly, making $\dRBH$ shallower or steeper pushes the range of black hole masses with relatively high event rates to higher or lower $M_{\rm BH}$, respectively (Fig.~\ref{fig:dRBH}).  The values of $\alpha_{\rm BH}$ in the two panels of this figure differ by 1; the very sharp contrast in the shape and placement of their distributions with respect to $M_{\rm BH}$ demonstrates that a test of this sort gives a very sensitive measure of $\alpha_{\rm BH}$.  It is striking how a relatively modest change in the distribution of events with black hole mass can cause a dramatic shift to the slope of the apparent ``typical luminosity" (actually the region of greatest observed event rate) as a function of black hole mass, even transforming it from positive to negative. Thus, measurements of the TDE distribution could potentially yield inference of this parameter with an uncertainty
$\lesssim 0.5$.  When $\alpha_{\rm BH}$ is comparatively large, an effect noted in the previous subsection becomes important: more events with high $M_{\rm BH}$ lead to a larger fraction of partial disruptions because more and more events with a pericenter smaller than the critical value for full disruptions produce no flare at all as the star falls into the black hole before it can be disrupted.

\subsubsection{Effect of varying $\dNstar$}

The greatest effect of varying the stellar mass function is the relative proportions of high- and low-luminosity events, and this relative proportion depends primarily on the relative numbers of high- and low-mass stars.
Fig.~\ref{fig:dNstar} illustrates these trends by portraying the TDE distribution due to a pair of stellar mass functions chosen
to demonstrate the impact of the mass cut-off found in an old stellar population.
 One stellar mass function has the logarithmic slope of a Salpeter IMF up to a cut-off at $1 M_\odot$, representing an old stellar population dominated by low-mass stars. The other is a pure Salpeter IMF with no cut-off.
In both cases, $\alpha_{\rm BH} = 0$.  It is immediately obvious that concentrating the stellar population toward smaller masses also concentrates the event density toward lower luminosities, while concentrating it toward larger masses emphasizes high-luminosity events.  Although the black hole mass-dependence in the event rate is the same in these two examples, the greater incidence of high-mass stars in the no cut-off
case also permits more high-mass black holes to participate in disruptions. Just as for the black hole mass distribution, the counts of TDE rates with respect to peak luminosity can be a strong constraint on the shape of the stellar mass function, and therefore the history of star formation, in the cores of external galaxies.

\begin{figure*}
\centering
\raisebox{-7pt}[\height][0pt]{
 {  \includegraphics[width=0.45\linewidth]{sal.png}}
}
\hspace{0.02\linewidth}
 { \includegraphics[width=0.49
 \linewidth]{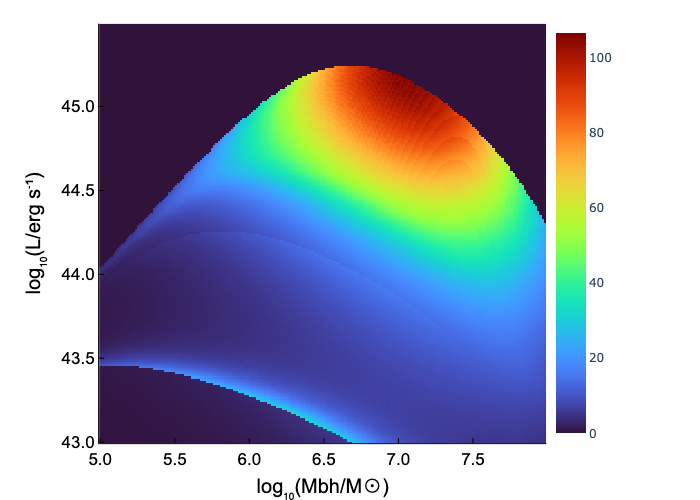}}
\caption{The total (full + partial) event rate density in the $\log M_{\rm BH} \times \log L_{\rm peak}$ plane for flux-limited samples: 
(left) using the single power-law main-sequence mass-radius relation; (right) the broken power-law main-sequence mass-radius relation.  In both cases, the stellar mass function is a Salpeter IMF with no mass cut-off, and $\alpha_{\rm BH} = 0$.  Note that the right plot extends to higher luminosities than the left plot.
Rates (arbitrary units) are shown on a linear scale calibrated by the colorbar.
}
\label{fig:mass-radius}
\end{figure*}

\subsection{Effects of uncertainties in stellar physics}

{As we have already pointed out, the peak luminosity is quite sensitive to $R_*$, varying $\propto R_*^{-5/2}$.  Thus, uncertainty in the main-sequence mass-radius relation for high-mass stars creates uncertainty in the maximum luminosity achievable in a TDE.   The number of such events that could be observed is $\propto L^{3/2}$, so this number depends even more strongly on the stellar radius, $\propto R_*^{-15/4}$.}

To show this effect clearly, the two panels of Fig.~\ref{fig:mass-radius} contrast the prediction made with our single power-law mass-radius relation with the prediction of the broken power-law relation.  For both cases, the stellar mass function is a Salpeter IMF, i.e., $dN_*/dM_* \propto M_*^{-2.35}$.
The event rate dependence on black hole mass $d{\cal R}_{\rm BH}/dM_{\rm BH}$ is independent of $M_{\rm BH}$. Despite the exact match in the stellar mass functions and the black hole mass-dependence, the distributions in the 
($M_{\rm BH},L_{\rm peak})$ plane are quite different.  The more precise mass-radius relation stretches the luminosity range of the region of high probability both to low and high luminosity, but particularly on the high end.

\section{Population Estimations}
\label{sec:population}

The question now arises as to how well, under the terms of our model, the observed population can be reproduced, and how tightly the observed population constrains the two mass functions.  Before attempting to answer this question, we will describe the available data.

\subsection{Observational samples}
\label{sec:obs_sample}

\subsubsection{Generic characteristics of TDE surveys}

As we have already mentioned, flux-limited samples are always biased toward high-luminosity events because the volume within which they can be seen is $\propto L^{3/2}$. A second bias is unique to TDEs. It arises  at low luminosities, for which it can be hard to pick out a flaring point-source against the integrated stellar light within the photometric aperture.  Unfortunately, it is much more difficult to evaluate the extent to which this affects detectability than the accessible volume effect.  Whether a flare of a given bolometric luminosity can stand out against background light depends on the size of the photometric aperture, the distribution of galactic core surface brightnesses, the distances of the events, the bolometric correction for the band of observation, and the sensitivity of the survey.

A simple estimate suggests that this effect may act at a significant level to suppress discovery of lower-luminosity TDEs.  At redshift $z = 0.05$, which is fairly typical for known TDEs, $1^{\prime\prime}$ corresponds to a physical distance of 1~kpc for $H_0 = 70$~km~s$^{-1}$~Mpc$^{-1}$.  As shown by \citet{French+2020} for the hosts of four TDEs, as well as for a number of post-starburst and early-type-galaxies, the $g$-band surface brightnesses of many galaxies are in the neighborhood of 19~mag at $\approx 1$~kpc from their centers.  For an effective aperture of $\pi$ square arc-seconds, such a surface brightness corresponds to $\nu L_\nu^{(g)} \approx 3 \times 10^{42}$~erg~s$^{-1}$.  If surveys can detect only flares that increase the flux from the host galaxy's center by at least $10\%$, flares with $\nu L_\nu^{(g)} < 3 \times 10^{41}$~erg~s$^{-1}$ would be lost. The bolometric correction for $g$-band luminosity in TDEs is not well known, but could be as much as $\sim 50$ if the spectrum is exactly thermal with a temperature of $3 \times 10^4$~K.  Thus, the bolometric luminosity cut-off could be at a level as high as $\sim 1 \times 10^{43}$~erg~s$^{-1}$.

\subsubsection{The observational sample}

We use the 
\texttt{ManyTDE} sample \citep{MvV},
which contains 67 events.
{This sample 
combines the events reported in three papers: \cite{vanVelzen+2019,Hammerstein+2023,Yao+2023},
with several additional ones discovered more recently.} 
For the work we present here, we use their tabulated peak bolometric luminosities and black hole masses determined from the galactic masses.  The measurement errors for the luminosities have a log-normal distribution that {we estimate has width $\sigma_{\log L} = 0.1$, largely due to the systematic error in bolometric corrections to the measured g-band luminosity.}
The black hole masses were estimated using the correlation between host-galaxy stellar mass and dynamically measured black hole masses presented in \citep{Greene2020}. These estimates also have systematic error as well as random error. For example, alternative estimates using the $M_{\rm BH}-\sigma$ relation from \citet{Greene2020}, when available, often differ by a factor of 10.  In addition, the data on which the $M_{\rm BH} - M_{\rm gal}$ correlation is based has only a handful of examples with $M_{\rm BH} \lesssim 1 \times 10^7 M_\odot$.
To take account of these uncertainties we suppose that they, too, can be described by a log-normal distribution, but with width {$\sigma_{\log M} = 0.5$}.
For both quantities, we construct a ``smoothed" probability density by convolving the nominal probability density with the associated log-normal distribution.

It is also interesting to note that the \texttt{ManyTDE} sample exhibits plentiful numbers of events in the range $1 \times 10^{43} - 1 \times 10^{44}$~erg~s$^{-1}$, and a few up to a few $\times 10^{45}$~erg~s$^{-1}$, but none of its 67 events has a luminosity $< 1 \times 10^{43}$~erg~s$^{-1}$.  This sharp low-luminosity cut-off in the sample suggests that the surface brightness effect we described in the previous subsection may be acting.  Both because there are no examples of such low luminosity in the observed sample and because their lack may be due to a selection effect, we will not display predictions for $L < 1 \times 10^{43}$~erg/s.

Lastly, as illustrated in Fig.~\ref{fig:obsdata}, the distibution of events in the  ($M_{\rm BH},L_{\rm peak})$ plane is quite broad and does not at all follow any sort of ``main sequence".  If one treats separately the cases with $M_{\rm BH} > 1 \times 10^8 M_\odot$, a separation we have already argued is demanded by the dramatic change in dynamics for events with such large black holes, the data display no obvious trends.
The scatter is so large that, to test models against TDE population data, it is essential to treat the data as a fully 2D distribution, rather than searching for mean trends (e.g., as attempted without excising the $M_{\rm BH} > 10^8 M_\odot$ events in \cite{Mummery+2025}).

\subsubsection{Our sub-sample }

As explained in Sec.~\ref{sec:model}, our model, like any model designed to treat ordinary TDEs rather than extreme TDEs, is not suitable for   $\MBH >10^8 M_\odot$.
Hence, we exclude from our sample the 8 TDEs whose estimated SMBH mass is in this range. Notably, almost all of these events are also high luminosity events, which would be quite rare in a volumetric sample.

\begin{figure*}
\includegraphics[width=1.1\linewidth]{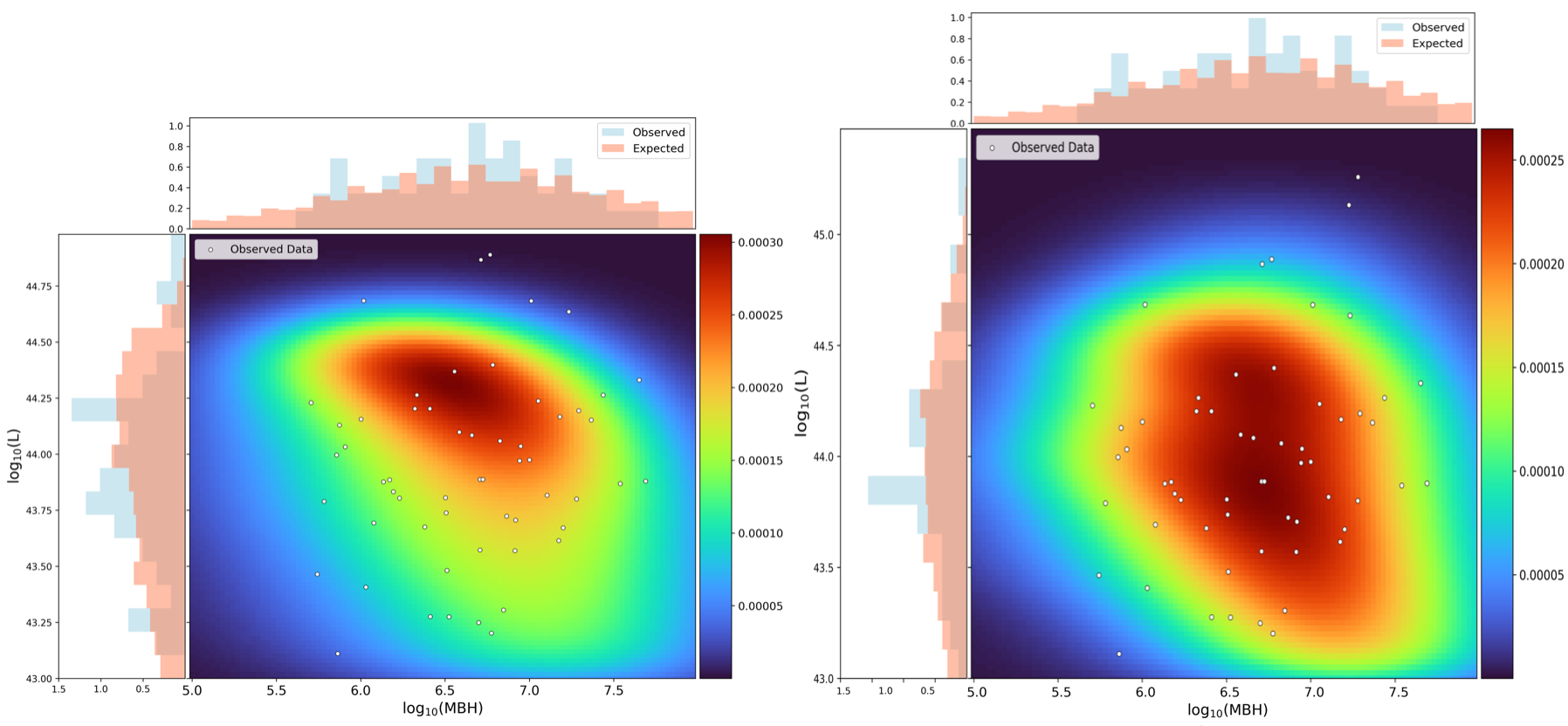}
\caption{The probability distribution function (PDF) for the best-fit parameters when we use the original MAMS mass-radius relation and the sub-sample of 57 data points (left) and the revised mass-radius relation with the sample of 59 data points (right).  Note the wider range of luminosities shown in the right panel. To account for observational errors, we smooth the PDF with Gaussians of width $\sigma_{\log_{10}(\MBH)}=0.5$ and  $\sigma_{\log_{10}(L)}=0.1$. The value of the PDF is represented by color contours calibrated by the linear color scale shown to the right of the plot. 
The observed data points of our sub-sample (without error bars) are marked by small circles. The histogram on top of the color contour plots shows the one-dimensional distribution of black hole masses, while the one on the left of the plot gives the one-dimensional distribution of the luminosities. 
}
\label{fig:pdf}
\end{figure*}
 
At first, we also exclude 
2 additional events whose estimated $\MBH$ is less than $10^8 M_\odot$ but whose observed luminosity is larger than $ 10^{45}$~erg~s$^{-1}$.
As explained earlier, $10^{45}$~erg~s$^{-1}$ is an upper limit to the luminosity that can be obtained with a single power-law mass radius relation when the maximum stellar mass is $30M_\odot$.  Lying just above this limit, these two events would have a disproportionate impact on likelihood statistics. This leaves us with a sample of 57 events. However, when we later consider the revised mass-radius relation, we reinstate these 2, so we include all events that satisfy $\MBH < 10^8 M_\odot$.

\subsection{MCMC}

Using the ``smoothed" probability density described earlier, we estimate the likelihood of an observed data set $[L_{{\rm peak},i} ,M_{{\rm BH},i}]$ as:
\begin{equation}
{\cal L} (\tilde \theta) = \prod_{i=1}^N P(L_{{\rm peak},i}, M_{{\rm BH}_i};\tilde \theta)
\label{eq:Lik}
\end{equation}
where $\tilde \theta \equiv (M_b, \alpha_1, \alpha_2, \alpha_{\rm BH})$ denotes the parameters of the model population. For estimating $L_{\rm peak}(M_*,\MBH)$ we use the canonical parameters given in \cite{Krolik+2025}. Thus, 
in the statistical inferences we report, {\it no} parameter in our physical model for the peak luminosity in a TDE flare is varied. 
We vary only the parameters of the stellar and black hole distributions. 

\begin{figure*}
\includegraphics[width=0.49\linewidth]{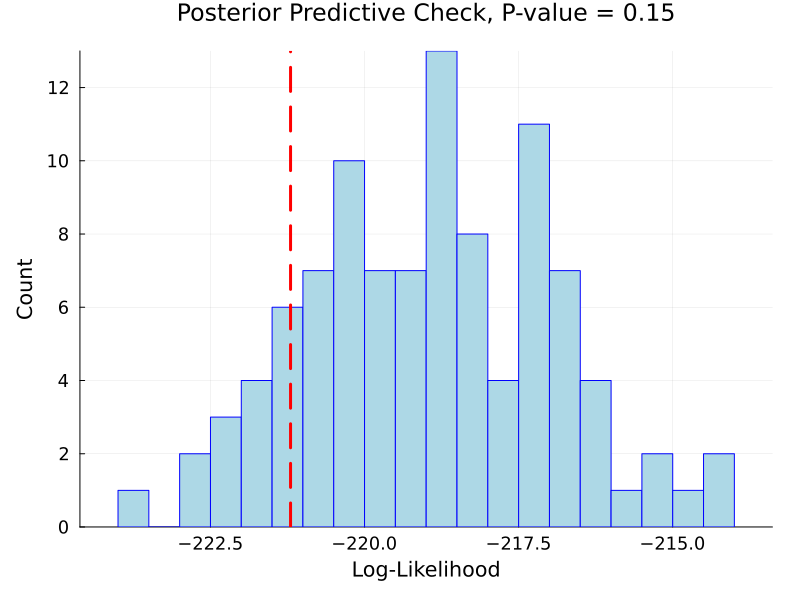}
\includegraphics[width=0.49\linewidth]{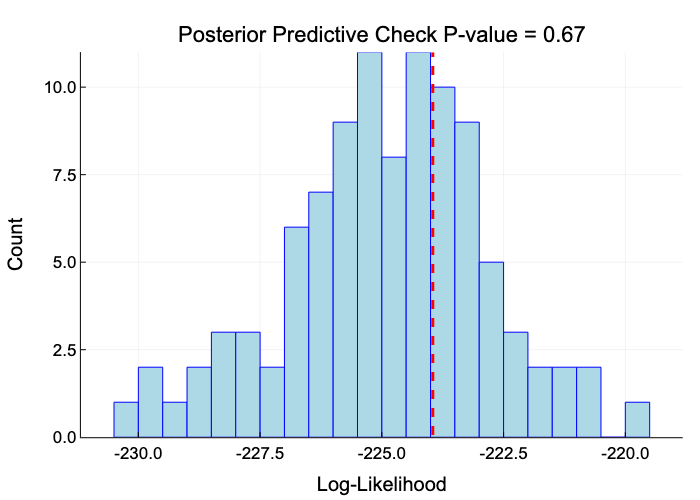}
\caption{The distribution of likelihoods for the best-fit parameters derived from 100 synthetic datasets constructed in the $p$-value procedure (blue histogram) compared to the likelihood of the actual data's best-fit parameters (red vertical line). For the single power-law mass radius relation (left) and  for the modified mass-radius relation (right). 
The likelihood of the observed data is better than the likelihood obtained in 15\% of the synthetic data sets for the single power-law model and is better than obtained in 67\%  of the synthetic data sets for the  modified mass-radius relation. 
}
\label{fig:likelihood_distribution}
\end{figure*}

\begin{figure*}
\includegraphics[width=1.0\linewidth]{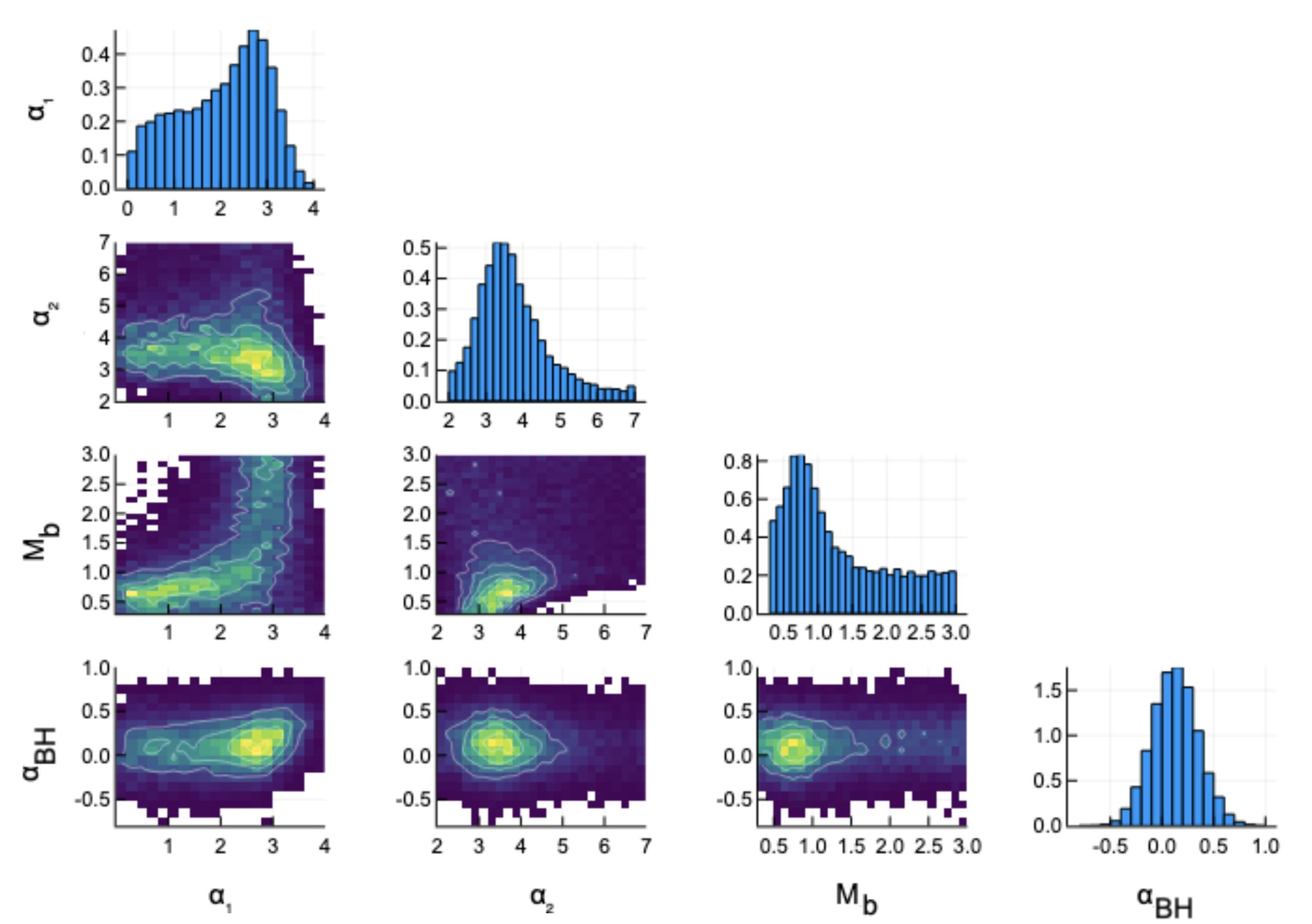}
\caption{A triangle diagram illustrating the parameter distribution for the 57 TDE sample when a single power-law mass-radius relation is used.  The underlying MCMC procedure took 50000 steps.
The color plots show the two-dimensional marginalized distribution of different pairs of parameters; the histograms show the one-dimensional marginalized distributions.
Note the independence of $\alpha_{\rm BH}$ from all other parameters and the weak positive correlations between 
$M_b$ and $\alpha_1$ and $\alpha_2$. These correlations ensure that even when $M_b$ is larger, the population is dominated by low mass stars.
}
\label{fig:MCMC}
\end{figure*}

We performed a
Bayesian inference on the set of  population parameters, $\tilde \theta$, by comparing the simulated joint
distribution of  $(L_{\rm peak},M_{\rm BH})|_{\tilde \theta}$  to the observational data.
Using a Metropolis--Hastings sampler, we find both $\hat\theta$, the most likely set of values of the parameters, and the distribution of parameters around it.  
Lacking any other information, our priors are  uniform within the ranges: $0.3 M_\odot < M_b < 3 M_\odot$, $1.0 < \alpha_1 < 4$, $3< \alpha_2< 7$ and $-1 < \alpha_{\rm BH} < 1 $. 
The parameter range for the stellar distribution function includes those associated with both the Salpeter and Kroupa IMFs. The range of
all the other parameters is large enough to let the Markov chain Monte Carlo (MCMC) procedure find the best fit without any imposed bias.
The limits on  $\alpha_2$ \ allow for cut-offs of varying sharpness, as would be the case for an old stellar population. 
Initially, we included values of $\alpha_2 < 3$ and a wider range of $\alpha_{\rm BH}$, but quickly found that the likelihoods were always very small, making it unworthwhile to expend the computer time exploring them.

Once we obtain the set of best fit parameters $\hat \theta$ and its probability density function (PDF), $\hat p([L_{{\rm peak},i},M_{{\rm BH},i}],\hat \theta)$, we test the validity of the model 
using the posterior predictive $p$-value method, also called the Bayesian $p$-value procedure. We first generate 100 synthetic datasets from the posterior predictive distribution (using the best-fit parameters of the MCMC sampling).
For each synthetic dataset, we then carry out a MCMC procedure to find its best-fit parameters and the corresponding PDF.  After calculating the test statistic (log-likelihood) for each synthetic dataset, we compare the results to the observed test statistic. 
The $p$-value is the proportion of synthetic datasets with test statistics poorer than for the observed data.  In other words, assuming that the best-fit model based on the observed data is correct, we compare the likelihood of the observed data to the likelihood of synthetic data created as realizations of our best-fit model.  To the degree that the actual data do the best (large $p$-value), we gain confidence in the model.

\subsection{Results}

We present two sets of results generated using this method,
one using our original single power-law fit to the MAMS mass-radius relation, the other using a broken power-law mass-radius relation that more accurately captures the actual function $R_*(M_*)$. Thus, employing it does {\it not} inject a new adjustable parameter; rather, it improves our account of stellar physics within our physical model for TDEs. As already remarked, the first analysis uses the subsample with 57 events, the second restores two high-luminosity events with $M_{\rm BH} < 10^8 M_\odot$, enlarging the sample to 59.

\subsubsection{{Quality of fit}}

The plot in  Fig.~\ref{fig:pdf} (left) shows the key results from the first analysis, the one with the single power-law approximation to the mass-radius relation. 
It compares the PDF computed using the most likely parameters (see Table~ \ref{tab:results}) with the observed points of the  corresponding sub-sample.   The match between the two is clearly quite good.   Its quality can be quantified in several ways.  One measure is given by comparing the one-dimensional distributions of the predicted black hole mass and peak luminosity to the observed distributions.  Modulo local fluctuations, both distributions are quite well reproduced.
Another quality test is shown in Fig.~\ref{fig:likelihood_distribution}, depicting the results of the posterior predictive estimates. 
The predictive $p$-value is 0.15, i.e., the resulting likelihood is better than that of 15\% of the 100 synthetic data sets sampled. 

\subsubsection{Determination of the parameters defining $\dNstar$ and $\dRBH$}

Fig.~\ref{fig:MCMC} and Table~\ref{tab:results}
show the best-fit parameters and give measures of how well they are determined.
The former shows the ``triangle diagram" for the two-dimensional and one-dimensional marginal distributions of  the four parameters;
the latter gives the values for which the greatest  likelihood is obtained, the average values of the parameters, and their standard deviations.

Two clear statements about the stellar mass and black hole populations stand out from this data:
\\ $\bullet $ Our physical TDE model requires a value of the
parameter $\alpha_{\rm BH}$, (bottom row) within $\pm 0.25$ of 0.
In other words, the event rate per star ${\partial {\cal R}_{\rm BH}}/\partial M_{\rm BH}$ depends very weakly on $M_{\rm BH}$.  This quantity is the product of the black hole mass function $dN_{\rm BH}/dM_{\rm BH}$ and the propensity for triggering TDEs as a function of $M_{\rm BH}$.
This narrow permitted range for $\alpha_{\rm BH}$ is essentially independent of all three stellar mass function parameters.
\\ $\bullet $ The likelihood that stars with mass $\gtrsim 2 M_\odot$ are present in sizable numbers is very small: $M_b$ is unlikely to be $\gtrsim 2 M_\odot$, and $\alpha_2$ is almost certainly $\gtrsim 3$.
In other words, the population of disrupted stars is rather old, with at most a very small number of massive stars.

Although the constraints on the stellar mass distribution yield a clear qualitative statement, their specific values are at most constrained by loose bounds.  There is little to distinguish any of the values of
$\alpha_1$ from 1 to 3.
This range includes both the Kroupa and the Salpeter slopes for the low-mass IMF.
The high-mass slope $\alpha_2$
could be anywhere from 3 to 6, and all these values give the stellar mass function for high masses a steep slope.  Although the nominally best-fit cut-off mass $M_b$ is
$\simeq 0.8M_\odot$, it could be as much as $1.5M_\odot$, but it cannot be any higher.

\begin{table}
    \centering
    \begin{tabular}{|c||c|c|c|}
      \hline
         $\quad\quad\quad$& Best & Average & $\sigma$ \\   \hline\hline
        $M_b$ & 0.61 (0.78) & 1.5 (1.0) & 0.79  (0.41)\\   \hline
        $\alpha_1$ & 1.2 (1.2) & 2.3 (1.7)  & 0.65 (0.51) \\   \hline
        $\alpha_2$ & 3.6 (4.8)  & 4.1 (4.8)   & 0.9 (0.68)   \\   \hline
        $\alpha_{\rm BH}$ & 0.023 (-0.05)  & 0.15 (0.134)  & 0.22 (0.221)\\   \hline
    \end{tabular}
    \caption{Best-fit parameters of the stellar and black hole distributions. The values in parentheses correspond to the solution without an upper luminosity cutoff and with a luminosity estimate based on the revised mass-radius relation. The average values and the standard deviations (denoted by $\sigma$) are for the one-dimensional marginalized distributions.}
    \label{tab:results}
    \vskip -0.5cm
\end{table}

Weak correlations exist between $M_{b}$ and $\alpha_1$, as well as between  $\alpha_2$ and both $\alpha_1$ and $\alpha_2$.  The sense of these correlations is for the slopes to increase with $M_b$. This is consistent with the tendency of the solution to prefer low mass stars. To compensate for a higher cut-off mass, steeper stellar mass function slopes are selected.

\subsection{ A modified stellar mass-radius relation}

As we have discussed already, the population predicted when a better approximation to the mass-radius relation is used can change significantly (see Fig.~\ref{fig:mass-radius}).
The more gradual increase of $R_*$ with $M_*$ for higher-mass stars increases $L_{\rm peak}$, permitting events with higher luminosity, while the slightly steeper slope of $R_*(M_*)$ for low-mass stars leads to somewhat higher luminosities for the smallest stars.

When the same MCMC procedure is repeated using the modified mass-radius relation, the results are very similar, but better.   The two high-luminosity events with $M_{\rm BH} < 1 \times 10^8 M_\odot$ that were previously excluded can now be explained, and the overall shape of the distribution matches the shape of the observed distribution even more closely.
Just as we did for the original mass-radius case, we compare the best-fit PDF for the improved mass-radius relation with the observed data in Fig.~\ref{fig:pdf} (right panel).  Five events that lay outside the region with significant probability as predicted with the original mass-radius relation are now in a region with probability density $\simeq 1/2 - 2/3$ the maximum density.  In addition, roughly half of all the events have $L_{\rm peak} \lesssim 1 \times 10^{44}$, providing a better match to the observed distribution easily visible in the figure.  The probability density for this lower-luminosity region when the cruder mass-radius approximation was used was $\simeq 1/3 - 1/2 $ the maximum, but increased to $\simeq 1/3 - 9/10$ the maximum with the improved version.

More quantitatively, the posterior predictive $p$-value=0.67 is shown in  Fig.~\ref{fig:likelihood_distribution}. In other words, the likelihood of the real data is better than that of $67\%$ of the synthetic data cases. Thus, although both the single power-law and the two power-law mass-radius relations give statistically acceptable fits to the data, it is clear that the revised model is preferable, both by accounting for the higher luminosity events and also giving a better $p$-value for the posterior predictive fit.  In addition to fitting better the higher luminosity values, the two power-law model produces a better fit to the data in the low $\MBH$, low $L$ region.

\begin{figure*}
\includegraphics[width=1.0\linewidth]
{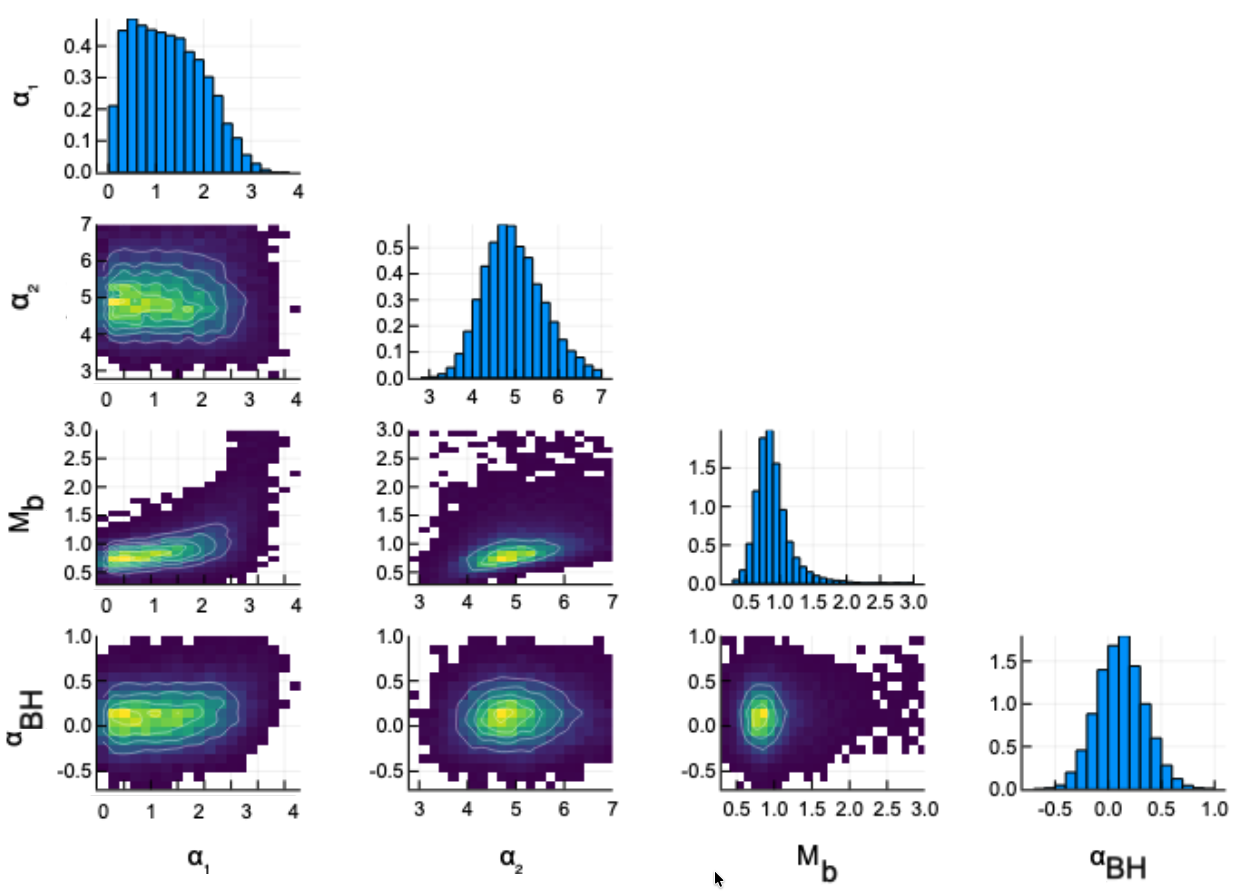}
\caption{A triangle diagram illustrating the parameter distribution for the 59 TDE sample when a dual power-law mass-radius relation is used. The underlying MCMC sampling had 50000 steps.   
The color plots show the two-dimensional marginalized distributions and the histograms show  the one-dimensional marginalized distribution of the different parameters.  Much like the parameter distributions based on the 57 TDE sample and a single power-law mass-radius relation, $\alpha_{\rm BH} \simeq 0$, nearly independent of all other parameters.  Similarly, the weak positive correlations between $M_b$ and $\alpha_{1,2}$ persist.  Relative to the other analysis, massive stars are somewhat more strongly suppressed: larger values of $\alpha_2$ are favored. }
\label{fig:MCMC_MR}
\end{figure*}

Even though the best fit parameters are slightly different (see Table~\ref{tab:results}), both are within the error bars of each other. A comparison of the two triangle diagrams (Figs.~\ref{fig:MCMC} and \ref{fig:MCMC_MR}) shows similar trends in both cases. The value of $\alpha_{\rm BH}$ remains very close to zero, independent of the stellar parameters.  The cut-off mass $M_b$ is still most likely to fall in the range $0.8 - 1.5M_\odot$, with very little chance it can be any higher.  The low-mass slope $\alpha_1$ is constrained to a slightly smaller range: $\simeq 1 - 2$ rather than $\simeq 1 - 3$.  Lastly, the new mass-radius relation favors a somewhat steeper cut-off slope: $\simeq 4 - 6$ rather than $\simeq 3 - 4$. 


\section{Discussion}

Our results have implications for a number of aspects of TDE dynamics and populations.

\subsection{Physical models for the luminosity}

Perhaps most importantly, recent years have seen a great deal of effort given to first-principles global simulations of TDEs.  These simulations begin with a star approaching a supermassive black hole and then follow the event all the way through the star's disruption and several orbits of its bound debris \citep{Ryu2023b,SteinbergStone2024,Price+2024,Abolmasov+2025}.  The consensus of these simulations is that during the first few $t_0$ after the disruption, {more than $99\%$ of} the bound debris mass occupies an irregular eccentric disk extending over a region $\gtrsim 1000r_g$.  Heat production in this disk is dominated by shocks within the flow.  A far smaller fraction of the bound debris (well less than 1\%) is deflected onto smaller orbits, on the scale of the star's pericenter.  These findings stand behind the functional relationship presented in eqn.~\ref{eq:Lpeak definition} that predicts the peak optical/UV luminosity as a function of the stellar and black hole masses \citep{Krolik+2025}.  Additional luminosity in the soft X-ray band may be generated inside the pericenter, but the ability of outside observers to see it may be severely limited because the bulk of the debris is
optically thick as well as geometrically thick
\citep{Dai+2018,Krolik+2025}.
We regard the tight connection between realistic simulations of TDE dynamics and our model's prediction of the peak luminosity as a strength of our approach.

Previous models for the luminosity of TDEs \citep{Strubbe2009,MetzgerStone2016,Bonnerot+2021,
Dai+2018,Balbus-Mummery2018,Mummery-SB2020,Mummery+2024}, mostly developed before the global simulations were published, have generally 
assumed that within the first $\sim t_0$ after the disruption, a large part of the bound debris mass settles into a circular disk formed inside $\simeq 2r_p$ of the black hole, i.e., several tens of $r_g$.  This assumption differs substantially from the simulations' finding that the overwhelming majority of the bound debris remains much farther from the black hole ($\gtrsim 10^3 r_g$) for a considerable time.

Interestingly, the function $L_{\rm peak}(M_*,M_{\rm BH})$ with which we work (eqn.~\ref{fig:Lpeak} and \citet{Krolik+2025}) contains {\it no} free parameters. Its internal coefficients are instead calibrated from the simulations.  Consequently, the parameters that can be varied when predicting a TDE population have to do with the precision of the stellar mass-radius relation employed, the population of surrounding stars, the mass-distribution of supermassive black holes in galactic nuclei, and the propensity as a function of $M_{\rm BH}$ for a black hole to provoke tidal disruptions.

\subsection{Details important to TDE population model construction}

Several details of the comparison to the observed population are worth noting.  First, unlike any previous TDE population study, we have included the contribution of partial disruptions.  Although their comparatively low luminosity makes them a minority in flux-limited samples, they are not a small minority: for our best-fit parameters, they contribute $\sim 30\%$ of all detected events.  Consequently, all future discussions of the TDE population should incorporate them.

Second, as previous work has shown \citep{Ryu+2023}, the dynamics of disruptions undergo a sharp change when the star's pericenter is closer to the black hole's center than $\simeq 6 r_g$.   However, when $M_{\rm BH} > 1 \times 10^8 M_\odot$, the critical pericenter within which a full disruption occurs is smaller than this distance for virtually the entire range of stellar masses, excepting only the most massive  (this criterion is exact for Schwarzschild spacetimes but should apply in an angle-averaged fashion for Kerr spacetimes).  Thus, when $M_{\rm BH} \gtrsim 10^8 M_\odot$, unless the star is quite massive, {\it no} encounter with a supermassive black hole leads to its full disruption by the conventional pathway; it is either torn apart in a fashion very different from ordinary TDEs or it is captured directly into the black hole without suffering disruption outside the event horizon.

It follows that if a TDE is identified with a galaxy whose central black hole is known to be $\gtrsim 10^8 M_\odot$, {\it no} model for the peak luminosity of a disruption flare that applies to conventional disruptions should be applied to events in such a galaxy.   The TDE population for events with $M_{\rm BH} \gtrsim 10^8 M_\odot$ must therefore be described separately from events driven by smaller black holes.   The data themselves already hint at this fact: the distribution of $L_{\rm peak}$ for events with $M_{\rm BH} \gtrsim 1 \times 10^8 M_\odot$ has a shape very different from the distribution of the majority of TDEs (see Fig.~\ref{fig:obsdata}).

The converse of this argument is that it is possible extreme TDEs, those in which the pericenter is smaller than $\simeq 6 r_g$ but larger than the critical pericenter for direct capture, are the events appearing at $M_{\rm BH} \gtrsim 10^8 M_\odot$ and $L_{\rm peak} \gtrsim 1 - 3 \times 10^{45}$~erg~s$^{-1}$.   Unfortunately, at the present time there has been no trustworthy calculation of their peak luminosity, much less the time-dependence of their luminosity.  It is, however, intriguing that the luminosity range in question is $\sim 10^{-1} L_{\rm Edd}$ for $M_{\rm BH } \sim 10^8 M_\odot$.

\subsection{Statistical consistency between the hydrodynamics-based model of TDE flares and the observed population}

In this work, we have shown that a prediction of the peak luminosity based directly upon the simulation results reproduces the overwhelming majority of the observed population when combined with plausible distributions for the stellar mass and the black hole mass.  That such a match can be achieved is displayed visually in Fig.~\ref{fig:pdf} (left).  This figure demonstrates that there is very good overlap between the overwhelming majority of the observed events with $M_{\rm BH} < 10^8 M_\odot$ and the model prediction.  It is demonstrated in quantitative fashion in several ways.  First, Fig.~\ref{fig:pdf} (left) also shows that the distributions of events with respect to both $M_{\rm BH}$ and $L_{\rm peak}$ are likewise matched very well.  Second, the model's posterior predictive $p$-value is reasonably large, 0.15.

When a more realistic main-sequence mass-radius relation is substituted for a more approximate one, the match between the predicted probability distribution and the observed distribution becomes even better: the entire population of ``ordinary" TDEs (i.e., those with pericenters $> 6r_g$), not just a large majority, is well-fit.  Moreover, as illustrated in Fig.~\ref{fig:pdf} (right), the model distribution fits the observed distribution even more closely.  In addition to reproducing the highest luminosity events, it also provides a closer match to the shape of the distribution across the whole range of observed luminosities.  The latter effect is visible in the very close representation of the luminosity distribution.  These improvements combine to create an excellent $p$-value, 0.67. 

That a prediction grounded in detailed hydrodynamical simulations is consistent with observations provides evidence that our understanding of the fundamental physics of tidal disruptions is correct.

\subsection{The stellar population in galactic nuclei as inferred from this model}

Further support for this model comes from the fact that the parameters for which we find greatest likelihood describe a common situation: an old stellar population whose low-mass distribution is consistent with either a Salpeter or Kroupa IMF, although the modified mass-radius relation prefers Kroupa.  The key empirical fact driving this inference is that the density of TDEs as a function of peak luminosity---in a flux-limited sample---is fairly even between $\sim 1 \times 10^{43}$ and $\sim 5 \times 10^{44}$~erg~s$^{-1}$.   Because $L_{\rm peak}$ increases with $M_*$, and flux-limited samples are strongly biased toward higher luminosity events, such an even distribution can come only from a limited dynamic range in $M_*$.  Truncation of the stellar mass distribution due to age satisfies this criterion.  Another piece of evidence in favor of our expression for $L_{\rm peak}$ is that it produces the correct range of luminosities when the range of masses contributing runs from a few tenths of $M_\odot$ to $\sim 1 M_\odot$.

The same arguments explain why the most common mass of an observed disrupted star is $\simeq 1 M_\odot$.  The number of stars per $\log M_*$ that should appear in a flux-limited sample is $\propto (dN_*/d\log M_*) L^{3/2}$ because the differential mass function is a density and the accessible volume is $\propto L^{3/2}$.  This product peaks near $M_b$ when $M_b \sim 1 M_\odot$ because $\partial \Xi/\partial M_*$ is largest for $M_* \sim 1 M_\odot$.

In this context, it is important to recognize how sensitive flux-limited samples are to rare high-luminosity events.  Although the likelihood is greatest for fairly steep cut-offs in $dN_*/dM_*$ ($3 \lesssim \alpha_2 \lesssim 4$ for the simple mass-radius relation, $4 \lesssim \alpha_2 \lesssim 6$ for the broken power-law version), so the fraction of all stars with $M_* \gtrsim 3M_\odot$ is quite small, the fraction of observed events involving such stars ($\sim 10\%$) is orders of magnitude larger because $L_{\rm peak}$ increases with $M_*$ and the accessible volume is $\propto L^{3/2}$.

We also find that $\alpha_{\rm BH}$, the only parameter governing $\dRBH$, is tightly constrained to $0 \pm 0.25$.   To confirm this inference requires both better measurements of the supermassive black hole mass function and more complete theories of how the black hole mass influences its rate of driving TDEs.  It is entirely possible that a mass distribution for nuclear black holes that declines steeply with increasing $M_{\rm BH}$ combines with mechanisms by which more massive black holes are much more likely to disrupt stars in order to make $\dRBH$ flat in terms of $M_{\rm BH}$, or rising linearly in terms of $\log M_{\rm BH}$.

\subsection{A new, low-luminosity population of TDEs?}

To close, we remark that the general tendency for all stellar mass functions to peak in the range of low-mass stars implies that if we could compile 100\% complete {\it volume-limited} samples, there would be a great many more low-luminosity TDEs than appear in flux-limited samples.  
However, achieving 100\% completeness is made very dificult
by problems like distinguishing a low-luminosity TDE from an unresolved stellar background.   In fact, numerical experiments along this line suggest that the actual number per unit volume of TDEs with $L < 1 \times 10^{43}$~erg~s$^{-1}$ is likely to be at least $5 - 10 \times$ the number of higher luminosity examples.  The existence of such a large population of TDEs very difficult to detect
adds a new element to the comparison between theoretical predictions of TDE rates and observational measurements.

\section{Summary}

We have demonstrated that our optical TDE peak-luminosity model \citep{Krolik+2025} accurately predicts the bulk of the observed distribution in the ($M_{\rm BH},L_{\rm peak})$ plane. This model rests directly on detailed hydrodynamics simulations starting from before the disruption and running for several orbital periods of the debris \citep{Ryu2023b,SteinbergStone2024,Price+2024,Abolmasov+2025}.
There are {\it no} adjustable parameters in the physical model. 
The only quantities fit to the data are those describing the underlying populations: the stellar mass function and the black hole event-rate distribution. 

Before discussing our results we note that a comparison of the observed TDE population in the ($M_{\rm BH}, L_{\rm peak}$) plane to predictions made by models of light production in a TDE can be sensibly made only in terms of the full 2D distribution in this plane, and in conjunction with statements about the stellar mass distribution $dN_*/dM_*$ and the black hole event rate distribution $d{\cal R}_{\rm BH}/dM_{\rm BH}$.  The scatter in $L_{\rm peak}$ within the observed sample of events is so large that single-parameter ``trends" cannot capture the demographics of the population, and correlation analyses of trends with $M_{\rm BH}$ are sensitive to both the stellar mass and the black hole distributions.

Using MCMC sampling, we find that the data strongly favor a stellar population old enough that there are very few stars with $M_* \gtrsim 1.5$–$2 \,M_\odot$.
The current sample is not large enough to constrain tightly the slopes of the stellar function below and above the main-sequence cut-off mass other than to say that the high-mass slope must be much steeper.  Nevertheless, the allowed region for the low-mass slope includes the values for both of the most commonly used forms, the Salpeter and Kroupa IMFs.  We also find that, at fixed $M_*$, the volumetric event rate is nearly independent of black hole mass, $M_{\rm BH}$.

We have also demonstrated explicitly that partial disruptions must be included in any realistic statistical analysis of TDE samples. Although their peak luminosities are always lower than those of corresponding full disruptions, and they therefore contribute only a minority of events in flux-limited samples, they are unavoidable in physical loss-cone models, particularly when the loss cone is not full. Partial disruptions account for approximately $\sim 30\%$ of the total observed event rate in our modeling. Moreover, their distribution in both black hole mass $M_{\rm BH}$ and peak luminosity $L_{\rm peak}$ differs systematically from that of full disruptions. As a result, omitting partial disruptions would bias both the inferred population properties and the interpretation of observed TDE demographics.

Our model  suggests that there is a significant number of low-luminosity TDEs, whose peak luminosities fall below $10^{43}$ erg sec$^{-1}$. Such events are largely undetectable because of the intrinsic brightness of galactic centers that they must outshine. The rate of these events may be as high as five to ten times larger than the rate of detectable, brighter TDEs. This prediction may explain some possible discrepancies between the observed rate of TDEs and theoretical predictions.

Although our model explains the distribution of TDE peak luminosities and black hole masses for a large majority of the observed events, it is not applicable to a small minority taking place around very massive  ($\gtrsim 10^8 M_\odot$) black holes\footnote{In the current sample all these events have very high luminosity, implying that in a flux limited sample they are significantly over-represented compared to their volumetric rate. } . This limitation is expected and was explicitly anticipated in \citet{Krolik+2025}. For sufficiently massive black holes, the tidal radius ${\cal R}_T$ contracts to only a few $r_g$. When the pericenter distance is $\lesssim 6\,r_g$, strong apsidal precession forces the debris to make several turns around the black hole. These TDEs  must therefore be qualitatively different from ordinary ones (see Section~\ref{sec:model}). As a result, the functional dependence of the peak luminosity $L_{\rm peak}$ on stellar and black hole masses must differ fundamentally from that applicable to events with larger pericenter distances. Consequently, any statistical analysis of TDE samples spanning both $M_{\rm BH} \lesssim 10^8\,M_\odot$ and $M_{\rm BH} \gtrsim 10^8\,M_\odot$ must employ separate predictions for $L_{\rm peak}$ in the two regimes. In the absence of a predictive model for disruptions around higher-mass black holes, we therefore restrict the present analysis to events with $M_{\rm BH} < 10^8\,M_\odot$. It is important to note that some one-dimensional correlations between luminosity and black hole mass depend solely on this small group of high luminosity, high $\MBH$ events.

To conclude, our analysis points the way to a new method for constraining the nature of the stellar population in galactic nuclei.  With more data, the prediction made here, a predominantly old population with a small admixture of more massive, younger stars, could be significantly refined.  A population of this sort may be the result of more recent star formation taking place somewhat farther from the galactic nucleus, followed by a slow drift inward by the more massive stars \citep{Morris1993,Miralda2000,Freitag2006,Alexander2009,Linial2022,Balberg2024,Rom2025}.

\begin{acknowledgments}
JHK would like to thank Rosemary Wyse and Kevin Schlaufman for helpful conversations about the state-of-the-art in stellar population studies. 
This work was partially supported by NASA TCAN grant 80NSSC24K0100 (JHK), by the advanced ERC grant Multijets, and by the Simons SCEECS collaboration (TP).

\end{acknowledgments}

\begin{figure*}
\includegraphics[width=0.85\linewidth]
{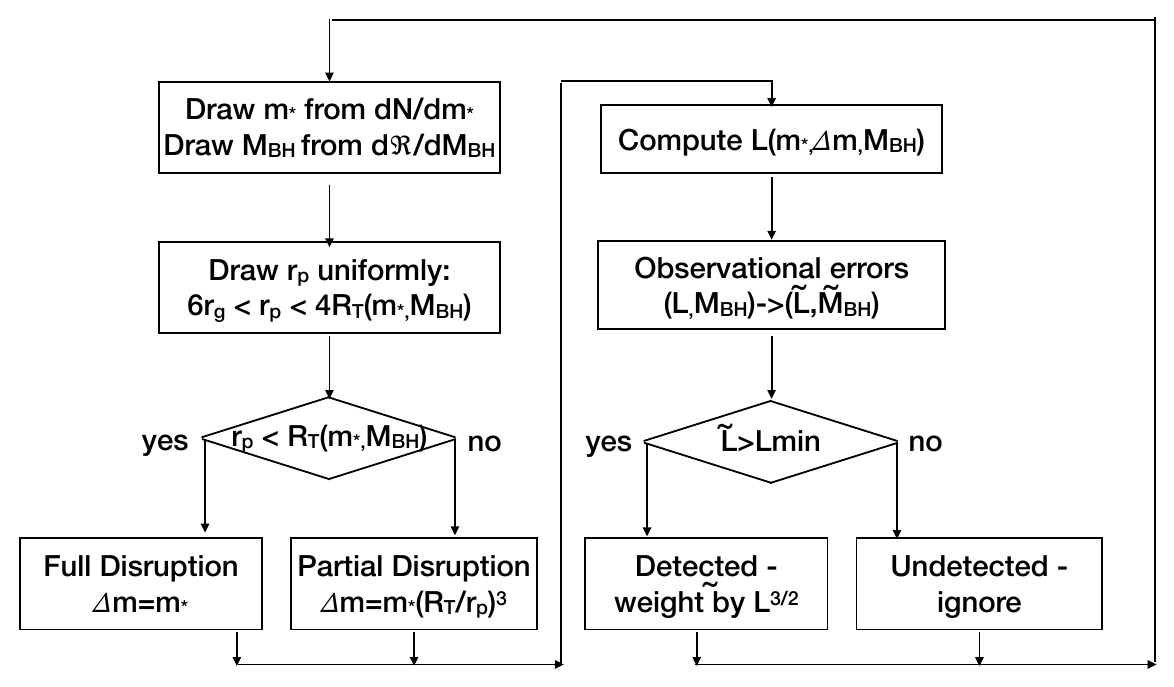}
\caption{The Monte Carlo loop.   
}
\label{fig:MCMCloop}
\end{figure*}

\vskip 0.75cm
\begin{center}
\uppercase{Appendix A: Monte Carlo Probability distribution function calculations}
\end{center}
\vskip 0.5cm

We calculate the probability distribution function using a Monte Carlo method, following these steps (see Fig.~\ref{fig:MCMCloop}):
\begin{itemize}
\item {\bf 1) Choose the stellar and black hole masses:} We pick randomly a value  of $\Ms$ from the  stellar mass distribution $dN/d\Ms$ 
and  a black hole mass $\MBH$ from the event rate-black hole mass distribution, $d{\cal R}/d\MBH$.

\item {\bf 2) Choose the pericenter distance:} We pick randomly a pericenter distance \footnote{Technically, one should use the square of the star's specific angular momentum, but this is the pericenter in the Newtonian limit.  For simplicity we use this approximation, but have checked that it is sufficiently accurate.} uniformly distributed within the range  $ 6 r_g < r_p< 4 R_{\rm T}(\ms,\MBH)$.\footnote{A transit at  $r_p \gtrsim 4R_{\rm T}$ doesn't result in any mass ejection. For $4 r_g < r_p < 6 r_g$, we expect an extreme TDE whose signature is very different from the optical TDEs that we are discussing, while $r_p < 4r_g$ results in a direct capture of the star. Both classes events are not included in our TDE population predictions.}  

\item {\bf 3) Distinguish full from partial disruptions:} The luminosity of events with  $6r_g <  r_p   $ \footnote{As discussed earlier events at $6r_g < r_p < 10 r_g $ might behave differently. However, lacking a better estimate we use Eq. \ref{eq:Lpeak definition} even for these events.  } 
depends on whether the disruption is full or partial, which in turn depends on $r_p/R_{\rm T}$.
\begin{itemize}
\item {\bf full disruption:} If $r_p < R_{\rm T}$, $\Delta M = \Ms$. 
\item {\bf partial disruption: } If $r_p > R_{\rm T}$,  $\Delta M/\Ms =  (R_{\rm T}/r_p)^3 $ \cite{Ryu+2020c}.
\end{itemize}

\item {\bf 4) Find the peak luminosity:} With $\Ms$, $\MBH$ and $x \equiv r_p/R_T$,   $L$  is calculated using Eqs. \ref{eq:Lpeak definition} or \ref{eq:partial}. 

\item {\bf 5) Account for observational errors:} To imitate observational errors, we smear the results over lognormal distributions. The peak luminosity $L$  is multiplied by a random factor with a lognormal distribution with characteristic width $\Delta \log L = 0.1  $, and the SMBH mass is similarly multiplied by a random factor with a lognormal distribution having a width $\Delta \log M_{\rm BH}= 0.5$. The resulting pair ($\tilde L, \tilde \MBH$) represent the ``observed" parameters of the event. 

\item  {\bf 6) Adjust for detectability:} The volume within  which an event of peak luminosity $L$ can be  detected is  $\propto L^{3/2}$, so we multiply each event's contribution by that factor and add it to the matrix of $p$ values. 
\end{itemize}

Repeating this process over a large number of random points we obtain a simulated 
binned probability density function (PDF) for the probability density of events: $p(L,\MBH,\tilde \theta)$, where $\tilde \theta \equiv (M_1, \alpha_1, \alpha_2, \gamma)$ denotes the parameters of the model.

\bibliographystyle{aasjournal}

\bibliography{biblio}

\end{document}